\documentclass[onecolumn]{emulateapj}

\usepackage{amsmath}
\def\snp{SN\,II-P} 
\def\snep{SNe\,II-P} 
\def\VmI{\hbox{$V\!-\!I$}} 
\def\fe{\ion{Fe}{2}}

\newcommand{\bvri}{\protect\hbox{$BV\!RI$} }

\begin{document}

\submitted{ApJ accepted December 2, 2008} 

\title {Improved Standardization of Type II-P Supernovae: Application to 
an Expanded Sample}
\shorttitle{Improved Standardization of SNe\,II-P}
\shortauthors{Poznanski et al.}

\author{Dovi Poznanski\altaffilmark{1},
Nathaniel Butler\altaffilmark{1,2}, 
Alexei V.\ Filippenko\altaffilmark{1},
Mohan Ganeshalingam\altaffilmark{1},
Weidong Li\altaffilmark{1},
Joshua S.\ Bloom\altaffilmark{1,3}, 
Ryan Chornock\altaffilmark{1}, 
Ryan J. Foley\altaffilmark{1,4,5}, 
Peter E. Nugent\altaffilmark{6}, 
Jeffrey M.\ Silverman\altaffilmark{1},
S. Bradley Cenko\altaffilmark{1},
Elinor L. Gates\altaffilmark{7},
Douglas C. Leonard\altaffilmark{8}, 
Adam A. Miller\altaffilmark{1}, 
Maryam Modjaz\altaffilmark{1}, 
Frank J. D. Serduke\altaffilmark{1}, 
Nathan Smith\altaffilmark{1}, 
Brandon J. Swift\altaffilmark{9}, and
Diane S. Wong\altaffilmark{1}}

\email{dovi@berkeley.edu}
\altaffiltext{1}{Department of Astronomy, University of California, Berkeley, CA
 94720-3411.}
\altaffiltext{2}{GLAST Fellow.}
\altaffiltext{3}{Sloan Research Fellow.}
\altaffiltext{4}{Harvard-Smithsonian Center for Astrophysics, 60 Garden 
Street, Cambridge, MA 02138.}
\altaffiltext{5}{Clay Fellow.}
\altaffiltext{6}{Lawrence Berkeley National Laboratory, 1 Cyclotron Road, 
Berkeley, CA 94720.}
\altaffiltext{7}{Lick Observatory, PO Box 85, Mount Hamilton, CA 95140.}
\altaffiltext{8}{Department of Astronomy, San Diego State University, Mail 
Code 1221, San Diego, CA 92182-1221.}
\altaffiltext{9}{Steward Observatory, University of Arizona, Tucson, AZ 85721.}

\begin{abstract}
	In the epoch of precise and accurate cosmology,
	cross-confirmation using a variety of cosmographic methods is
	paramount to circumvent systematic uncertainties. Owing to 
        progenitor histories and
	explosion physics differing from those of Type Ia SNe (SNe\,Ia), 
	Type II-plateau supernovae (\snep) are unlikely to be affected 
	by evolution in the same way. 
	Based on a new analysis of 17 \snep, 
	and on an improved methodology, we find that \snep\ are good 
	standardizable candles, almost comparable to SNe\,Ia. 
	We derive a tight Hubble diagram with a dispersion of 10\% in distance, 
	using the simple correlation between luminosity and photospheric 
        velocity introduced by
	\citet{hamuy02}. We show that the descendent method of
	\citet{nugent06} can be further simplified and that the correction for
	 dust extinction has low statistical impact.
	We find that our SN sample favors, on average, a very steep dust law with
	total to selective extinction $R_V<2$. Such an extinction law has
	been recently inferred for many SNe\,Ia.  
	Our results indicate that a distance measurement can be obtained 
	with a single spectrum of a \snp\ during the plateau phase combined with sparse
	photometric measurements.

\end{abstract}


\section{Introduction}

Compelling evidence for cosmic acceleration comes from distance
measurements to Type Ia supernovae \citep[SNe\,Ia; e.g.,][see
\citealt{filippenko05} for a review of earlier studies]
{astier06,wood-vasey07,riess07,kowalski08}. Yet even with unlimited
observational resources to improve statistical uncertainty, SN\,Ia
cosmology ultimately faces systematic uncertainties stemming from an
incomplete physical model of the phenomena.  Indeed, since the
physical conditions which give rise to the progenitors of the events
are extremely difficult to discern, with no real consensus having been
reached thus far, determining the nature of the evolution of SNe\,Ia
at high redshift remains a crucial challenge for precise and accurate
measurements of the fundamental cosmological parameters. While several
mechanisms have been devised to measure, test, and constrain SN\,Ia
systematics \citep[e.g.,][]{ellis08,foley08}, the possibility of
source redshift evolution warrants a continued exploration of
additional, complementary cosmographic methods.

A simple observation provides a compelling foundation for the use of
Type II-plateau supernovae (\snep) as standardizable candles: the
progenitors of several \snep\ have now been uncovered
\citep[e.g.,][and references therein]{li07,smartt08}, and all are
found to be red supergiants with somewhat similar initial stellar
masses (typically 8--16~M$_\odot$).  Consequently, the environments
and pre-SN evolution of such objects are reasonably tractable. Having
undergone core collapse and envelope ejection, these red supergiants
become SNe observationally defined by the presence of hydrogen in the
spectra and a ``plateau'' phase in their light curves
\citep[e.g.,][see \citealt{filippenko97} for a
review]{barbon79}. SNe\,II-P have been detected to redshift $z \approx
0.5$, and any evolutionary trends are likely to differ from those of
SNe\,Ia.

As in the case of SNe\,Ia, the intrinsic inhomogeneity in \snp\ peak
luminosity, and that produced by extrinsic factors such as dust, can
be calibrated.  There are two different approaches to determining
distance: the theory-based method, and the empirical
standardized-candle method.  The theoretical approach branches into
two: (1) the expanding photosphere method (EPM;
\citealt{kirshner75,eastman96}), a historical descendent of the
Baade-Wesselink method for variable stars \citep{baade26}, and (2) the
more modern ``synthetic spectral atmosphere fitting'' method
\citep[e.g.,][]{baron04,dessart06}.  Both techniques require the
comparison of high signal-to-noise ratio (S/N) photometry and spectra
of a given object to model atmospheres; for the second method,
synthesized spectra are computed for each SN in
detail. Distances based on improved versions of the theoretical
modeling for a handful of well-observed SNe\,II-P have been shown to
be precise to 10\%
\citep[e.g.,][]{leonard03,hamuy05,baron04,dessart06,dessart08}, not
greatly inferior to the best precision achievable with SNe\,Ia
\citep[7\%;][]{astier06}.  The requirement of high S/N observations,
however, limits the efficacy of this approach for events at larger
distances.

The standardized candle method first suggested by \citeauthor{hamuy02}
(\citeyear{hamuy02}; hereafter HP02) provides an independent empirical
way to achieve distances to \snep, and it is strongly anchored in
simple physics. In more luminous SNe, the H-recombination front is
maintained at higher velocities; the photosphere is farther out in
radius. When a \snp\ is on the plateau phase of its light curve (which
lasts for around 100 rest-frame days), there is a strong correlation
between the velocity of the weak \fe\ lines near $5000$\,\AA\ -- that
trace the photospheric velocity -- and the luminosity on the
plateau. Extinction corrections are based on a variety of different
methods. This empirical correlation was further studied by
\citeauthor{hamuy05} (\citeyear{hamuy05}; hereafter H05); application
to 24 SNe\,II-P in the Hubble flow yields a Hubble diagram in the $I$
band with a scatter of 15\% in distance.

\citeauthor{nugent06} (\citeyear{nugent06}; hereafter N06) modified
this technique by simultaneously combining both the extinction
correction (using the rest-frame $V-I$ color on the plateau) and the
\fe\ velocity correction to arrive at a simple correlation between
these parameters and luminosity. Given the smaller quantity of the
data required, the N06 method, or a variant thereof, is probably the
only framework that could be followed cost-effectively at high
redshifts.

The number of \snep\ for which distances have been derived, using any
available method, is $\sim$20. The paucity of such objects, and the
intractable biases that emerge from the way the sample was
constructed, warrants an analysis of more SNe, and push for improved
and better-tested correlations that will allow one to achieve the
precision and control of systematics that could make these SNe
competitive tools for precision cosmology.

In this work, we add 17 new objects to the current sample of low-$z$
\snep, almost doubling the size, and reanalyze 3 previously published
events.  We discuss the full sample construction in
\S\ref{s:sample}. In \S\ref{s:method} we present and modify the N06
fitting method. We discuss in \S\ref{s:results} the sample culling,
derive a new Hubble diagram, and find that the scatter is $\sim$10\%
in distance; we also consider the robustness and possible biases in
the method, and compare distances derived for SNe that occurred in the
same host galaxy, checking for internal consistency. Our conclusions
are summarized in \S\ref{s:conclude}.

\section{The Supernova Sample}\label{s:sample}

\subsection{The Previously Existing Samples} 

N06 compiled a list of 24 \snep, 6 of which had not yet been
published; one new event (SN 2004dh) was studied as part of the
Caltech Core Collapse Project (CCCP; \citealt{gal-yam07cccp}) and is
at low redshift, while the other 5 are at moderate redshifts,
$0.1<z<0.3$, found in the course of the Supernova Legacy Survey
\citep{astier06}. The other 18 SNe, previously collected by H05 and
HP02, were found and observed between 1986 and 2000, with a variety of
instruments. All of those SNe had $z<0.06$, though the majority were
at $z<0.01$. For such nearby objects peculiar velocities of the host
galaxies dominate the error budget, thus complicating the derivation
of robust correlations and distances.

Since a ``standard'' Hubble diagram of distance modulus vs. redshift
assumes that the sole contribution to the redshifts is cosmological,
we apply corrections to the heliocentric redshifts in order to account
for the peculiar velocities of nearby galaxies.  Following previous
authors, in the analysis below we correct the heliocentric redshifts
derived from the SNe and their host galaxies to the local Hubble flow
using the \citet{tonry00} prescription. We further assume a Hubble
constant $H_0=70$\,km\,s$^{-1}$Mpc$^{-1}$. Since we later fit for the
absolute magnitude of the SNe, $H_0$ merely serves as a convenient
scaling constant. For SN\,1999em, we derive a redshift using the
Hubble law and the Cepheid distance measured by \citet{leonard03},
$11.7 \pm 1$\,Mpc. We re-reduced the images of SN\,2004dh and report
here new photometry for that event. Finally, following H05, we remove
SN\,2000cb from the sample, since it is now believed to be a
SN\,1987A-like event rather than a classic \snp. Hereafter we will
refer to these 23 SNe as the ``HPN sample''(for Hamuy, Pinto, and
Nugent).

\subsection{The KAIT Sample: Nearly Doubling the Low-Redshift 
Census}\label{s:Ksample} 

The 0.76-m Katzman Automatic Imaging Telescope (KAIT) is mostly known
as the SN discovery engine of the Lick Observatory SN Search
\citep[LOSS;][]{filippenko01,filippenko05b}, but a substantial
fraction of its time is dedicated to follow-up broadband photometry of
nearby SNe. As part of this program, dozens of SNe\,II have been
monitored in the past ten years.  We select the \snep\ from the KAIT
sample, since other core-collapse SN subtypes are not necessarily
expected to obey the same correlations.  However, the \snp\ subclass
is not well defined. While historically SNe were classified by their
light-curve shapes, and SNe with a distinctive plateau were dubbed
\snep\ \citep[e.g.,][]{barbon79,doggett85}, it is unclear how constant
a plateau must be, how long that phase must last, and in which
photometric bands it appears. In fact, those SNe which show hydrogen
in their spectra but photometrically decline linearly (in magnitudes)
were called SNe II-L, but not a single one of the best-studied objects
(e.g., SN\,1979C) is considered a ``normal'' prototype (e.g,
\citealt{poznanski02}, and references therein; Poznanski et al. 2009,
in prep.).

With the modern emphasis shifting to spectroscopic follow-up
observations, the classification becomes, dishearteningly, even
muddier. It is often assumed by SN observers that only \snep\ exhibit
hydrogen lines having P-Cygni profiles (e.g., \citealt{schlegel96}),
but this claim has not been explored systematically. Lately, several
core-collapse SNe with extreme luminosities have been discovered
\citep[e.g.,][]{ofek07,smith07,smith08tf,quimby07,miller08,gezari08},
stretching the classification scheme beyond its current limits.

For the purpose of this work, we define the \snp\ subclass as follows:
(a) prominent hydrogen features in the spectra, (b) no narrow emission
lines indicative of interaction with the circumburst medium, and (c) a
prominent ``plateau'' phase in the $I$-band light curve.

\begin{deluxetable}
	{lrrrrrrl} 
	\tabletypesize{\scriptsize}
	
	
	\tablecaption{KAIT Sample \label{t:KAIT}} 
	\tablewidth{0pt} 
	\tablehead{ 
	\colhead{IAU Name} & 
	\colhead{Heliocentric $cz$}& 
	\colhead{Flow $cz$\tablenotemark{a}} & 
	\colhead{$t_{\rm expl}$\tablenotemark{b}} & 
	\colhead{$v_{\mathrm{Fe\,II} }$\tablenotemark{c}} & 
	\colhead{$m_I$\tablenotemark{d}} & 
	\colhead{$V-I$\tablenotemark{d}} & 
	\colhead{Discovery Reference} \\
	\colhead{ } &\colhead{(km\,s$^{-1}$)} &\colhead{(km\,s$^{-1}$)} &
	\colhead{ } &\colhead{(km\,s$^{-1}$)} &\colhead{(mag)} &\colhead{(mag)} } 
	\startdata 
	SN1999bg\tablenotemark{e} &  1275 &  1214(187) & 1259(5) &  4657(439) & 15.31(0.02) & 0.74(0.04) & \citet{li99a} \\ %
	SN1999em &   717 &   819( 70)\tablenotemark{f} & 1476(4) &  3382(171) & 13.23(0.03) & 0.58(0.03) & \citet{li99} \\ %
	SN1999gi &   592 &   643(187) & 1520(4) &  3697(200) & 14.00(0.02) & 0.86(0.03) & \citet{nakano99} \\ %
	SN2000bs &  8387 &  8387(187) & 1650(6) &  4115(276) & 17.67(0.09) & 0.51(0.10) & \citet{papenkova00} \\ %
	SN2000dc\tablenotemark{e} &  3117 &  3113(187) & 1762(4) &  4545(232) & 16.05(0.06) & 0.97(0.06) & \citet{yu00} \\ %
	SN2000dj &  4629 &  4625(187) & 1789(7) &  4784(250) & 17.00(0.09) & 0.59(0.11) & \citet{aazami00} \\ %
	SN 2001x &  1480 &  1329(187) & 1964(5) &  3800(203) & 14.74(0.02) & 0.57(0.03) & \citet{li01a} \\ %
	SN2001bq &  2623 &  2238(187) & 2034(6) &  4023(238) & 15.55(0.03) & 0.51(0.08) & \citet{nakano01} \\ %
	SN2001cm &  3412 &  3413(187) & 2064(0) &  4065(189) & 16.30(0.02) & 0.68(0.03) & \citet{jiang01} \\ %
	SN2001cy\tablenotemark{e} &  4478 &  4473(187) & 2086(6) &  3987(218) & 16.13(0.04) & 0.63(0.05) & \citet{ganeshalingam01} \\ %
	SN2001do\tablenotemark{e} &  3124 &  3121(187) & 2134(2) &  3927(311) & 15.75(0.03) & 0.97(0.05) & \citet{modjaz01b} \\ 
	SN2002gd\tablenotemark{e} &  2674 &  2418(187) & 2551(2) &  2774(315) & 16.78(0.03) & 0.74(0.05) & \citet{klotz02} \\ %
	SN2002hh &    48 &   297(187) & 2576(2) &  3692(435) & 13.68(0.05) & 2.67(0.05) & \citet{li02} \\ %
	SN2003hl &  2472 &  2105(187) & 2868(4) &  3561(293) & 15.60(0.02) & 0.98(0.04) & \citet{li03} \\ %
	SN2003iq &  2472 &  2105(187) & 2921(1) &  4043(179) & 15.27(0.03) & 0.66(0.03) & \citet{llapasset03} \\ %
	SN2004du\tablenotemark{e} &  5025 &  5020(187) & 3223(4) &  4855(302) & 16.60(0.04) & 0.53(0.07) & \citet{singer04} \\ %
	SN2004et &    48 &   297(187) & 3272(0) &  3980(194) & 11.39(0.01) & 0.52(0.01) & \citet{zwitter04} \\ %
	SN2005ay &   809 &   816(187) & 3453(4) &  3432(224) & 14.67(0.02) & 0.63(0.06) & \citet{rich05} \\ %
	SN2005cs &   463 &   505(187) & 3550(1)\tablenotemark{g} &  2102(211) & 13.98(0.02) & 0.69(0.03) & \citet{kloehr05} \\ %
\enddata

\tablenotetext{a}{Corrected velocity of recession using the local Hubble-flow 
model of \citet{tonry00}, unless otherwise noted.} 
\tablenotetext{b}{Explosion date, assumed to be the midpoint 
epoch (given as Julian Day $-$ 2,450,000) between 
the time of discovery and the last nondetection (the uncertainty being half this value), unless otherwise noted.} 
\tablenotetext{c}{Fit velocity of the Fe~II line at day 50, on the plateau.} 
\tablenotetext{d}{Measured at day 50, on the plateau.} 
\tablenotetext{e}{Not used in the final sample, see \S\ref{s:culling}.} 
\tablenotetext{f}{From the Cepheid distance of \citet{leonard03} and the Hubble law.} 
\tablenotetext{g}{From \citet{pastorello06}.} 
\end{deluxetable}

The light-curve shape criterion is intentionally defined here rather
loosely, so as not to preselect the best objects (beyond unavoidable
observational biases) and is revisited in detail in
\S\ref{s:culling}. Several dozen SNe in the KAIT sample pass these
criteria, but for 19 of them we have the data necessary for empirical
calibration and distance measurement; they constitute the main sample
analyzed here (Table \ref{t:KAIT}).  The \snep\ that were not selected
for this work were usually missing early-phase photometry to constrain
the explosion date, or lacked spectroscopic coverage during the
plateau, as required for measuring velocities. Two of our 19 chosen
SNe (SN\,1999em and SN\,1999gi) are already present in the HPN sample;
they are very well-studied objects
\citep[e.g.,][]{hamuy01,leonard02em,leonard02gi,dessart06}. We have
used them to ascertain that there are no systematic offsets between
our magnitude and velocity measurements and those of previous authors.
Thus, our sample consists of 17 new \snep.

The optical CCD images of the SNe were reduced as follows.
Flat-fielding and bias subtraction were performed automatically at the
telescope using calibration frames appropriate for the science
images. Galaxy subtraction and differential photometry were done using
the KAIT pipeline (Ganeshalingam et al. 2009, in prep.). Galaxy
template images were obtained with KAIT at least a few months after
the SN had faded beyond detection.  To ensure high-quality
subtractions, templates were acquired on photometric nights with
seeing $\le 3.0 ''$. Two independent routines were used to perform
galaxy subtraction. The first method is based on the ISIS package
\citep{alard98} as modified by Brian P. Schmidt for the High-$z$
Supernova Search Team \citep{schmidt98}. The second method is based on
the IRAF\footnote{IRAF: the Image Reduction and Analysis Facility is
distributed by the National Optical Astronomy Observatories, which is
operated by the Association of Universities for Research in Astronomy,
Inc. (AURA) under cooperative agreement with the National Science
Foundation (NSF).} task {\sc PSFMATCH} \citep{phillips95}. We
performed differential point-spread function (PSF) fitting photometry
to the results of both subtraction methods using the {\sc DAOPHOT}
package in IRAF to measure the SN flux relative to local standards in
the field. The results of the two subtraction methods were averaged.

Calibrations were obtained on photometric nights using both KAIT and
the 1~m Nickel telescope at Lick Observatory. \citet{landolt92}
standards were observed at a variety of airmasses on each night to
derive a photometric solution that we then applied to local standards
for each of the SN fields. Instrumental magnitudes were transformed to
standard Johnson-Cousins magnitudes using color terms derived from the
photometric solution of many Landolt standard-star calibrations (see
\citealt{modjaz01} and \citealt{li01cx} for more details). The
uncertainty in our subtraction and photometry pipeline is estimated by
injecting artificial stars with the same magnitude and PSF as the SN
into the original KAIT images and recovering them. The final
uncertainty is taken to be the scatter in recovering 20 artificial
stars added in quadrature with the calibration error.  We correct the
magnitudes for Galactic extinction using the maps of
\citet{schlegel98}.

We obtained optical spectra with the Kast double spectrograph
\citep{miller93} mounted on the Lick Observatory 3~m Shane telescope,
the Low Resolution Imaging Spectrometer \citep[LRIS;][]{oke95} mounted
on the 10~m Keck\,I telescope, and the Deep Imaging Multi-Object
Spectrograph \citep[DEIMOS;][]{faber03} on the 10~m Keck\,II
telescope. The position angle of the slit was generally aligned along
the parallactic angle to reduce differential light losses
\citep{filippenko82}.

\begin{figure}
	\epsscale{1}
\plotone{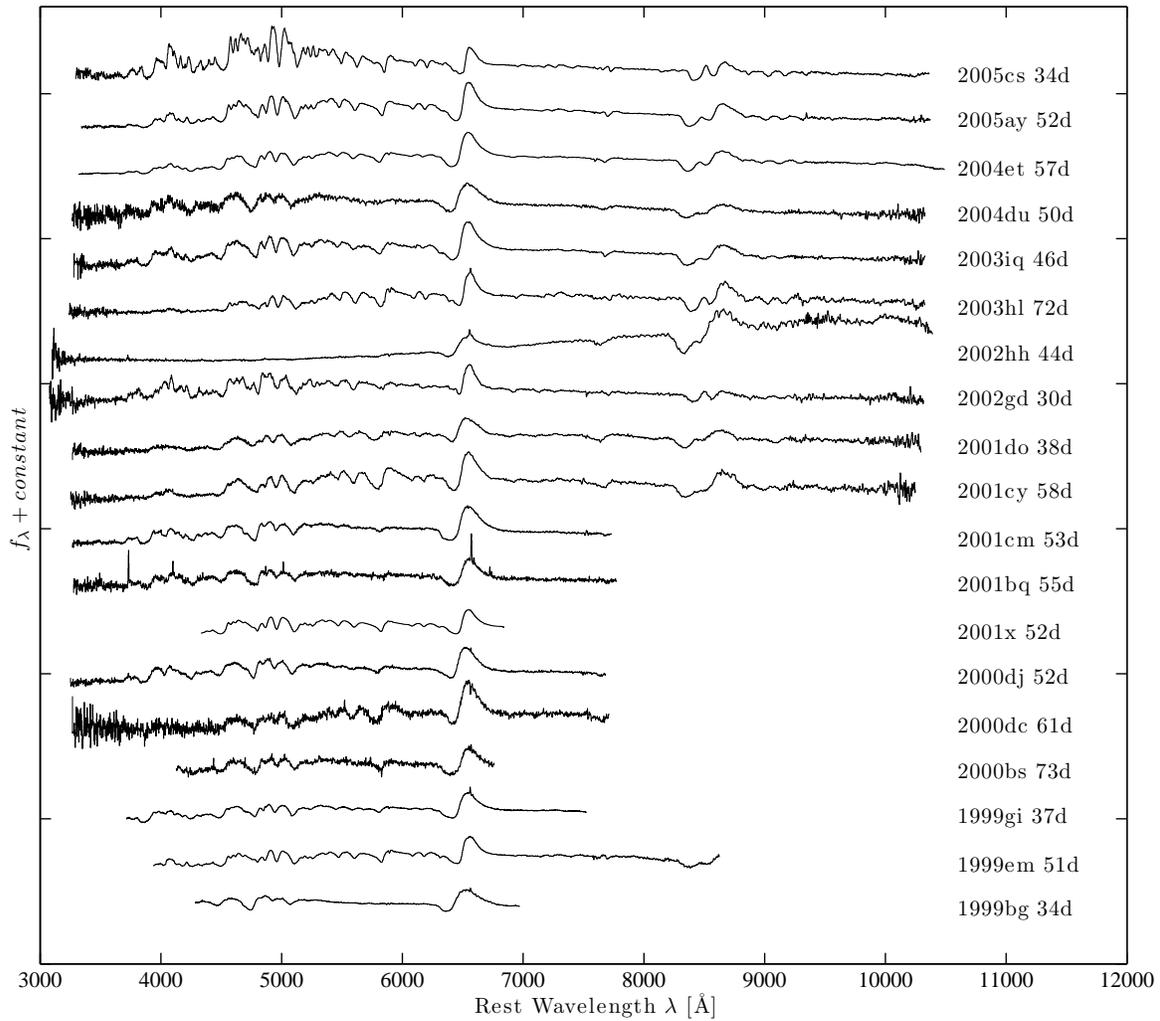}
\caption{A representative spectrum (the closest to day 50) for each of
the 19 SNe in our sample. The analysis herein makes use of about 75
spectra of these 19 events.\label{f:spec}}
\end{figure}

All spectra were reduced using standard techniques
\citep[e.g.,][]{foley03}. Routine CCD processing and spectrum
extraction for the Kast and LRIS data were completed with IRAF.  These
data were extracted with the optimal algorithm of \citet{horne86}.
CCD processing of the DEIMOS spectra was performed with a modified
version of the DEEP pipeline \citep[e.g.,][]{weiner05}.  This produced
rectified, sky-subtracted two-dimensional spectra from which
one-dimensional spectra were then extracted optimally \citep{horne86}.
The wavelength scale was derived from low-order polynomial fits to
calibration-lamp spectra.  Small wavelength shifts were then applied
to the data after cross-correlating a template night-sky spectrum to
the sky spectrum extracted near the SN position on the slit.  Using
custom routines, we fit spectrophotometric standard-star spectra to
the data in order to flux calibrate the spectra and to remove telluric
lines \citep{wade88, matheson00}. Information
regarding both our photometric and spectroscopic data (such as
observing conditions, instrument, data reducer, etc.) was obtained
from our SN database (SNDB).  The SNDB uses the popular open-source
software stack known as LAMP: the Linux operating system, the Apache
webserver, the MySQL relational database management system, and the
PHP server-side scripting language  (Silverman et al. 2009, in prep.).
Figure \ref{f:spec} shows one
spectrum of each of the 19 chosen SNe.

\begin{figure}
	\epsscale{0.7}
\plotone{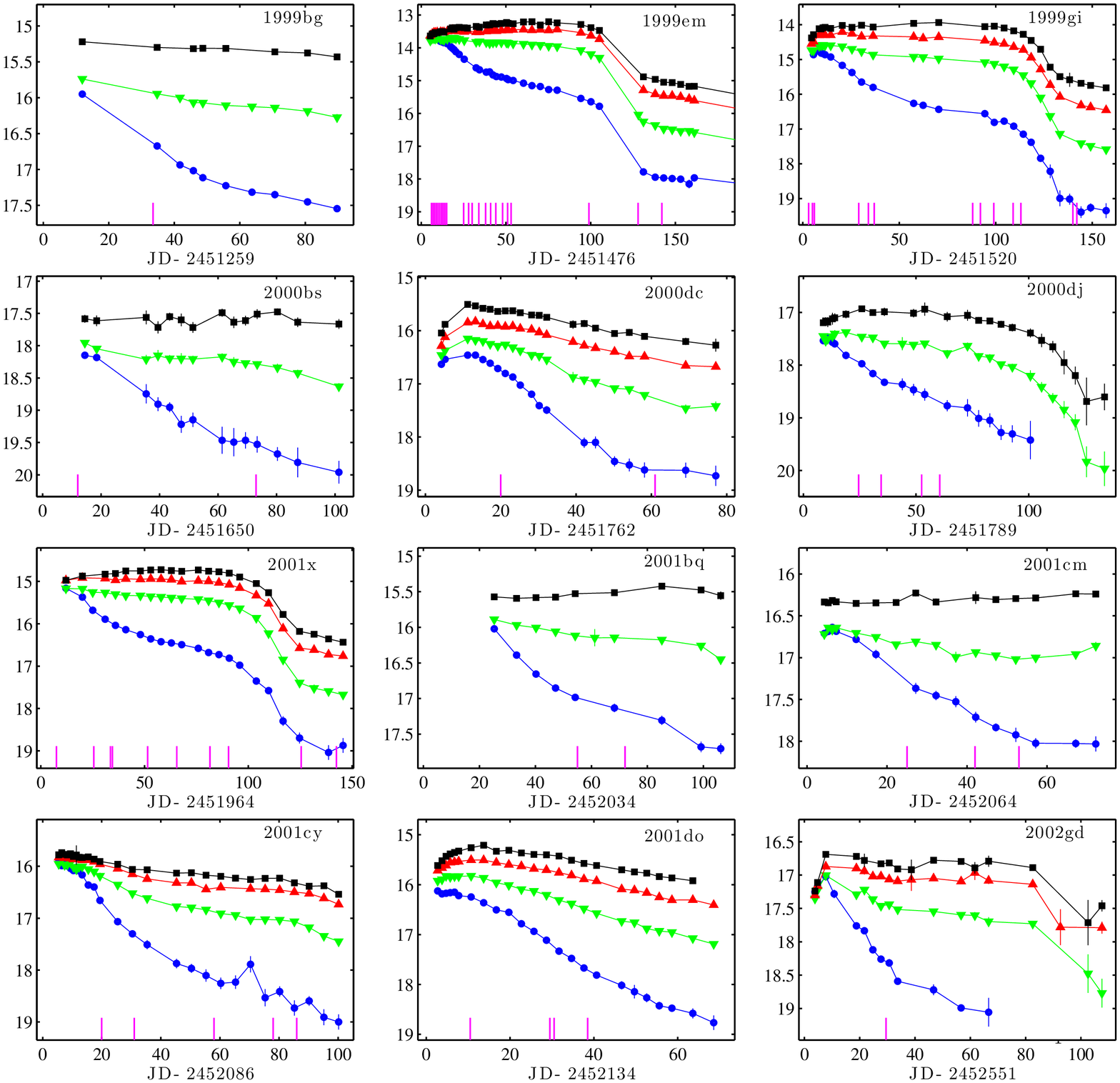}
\plotone{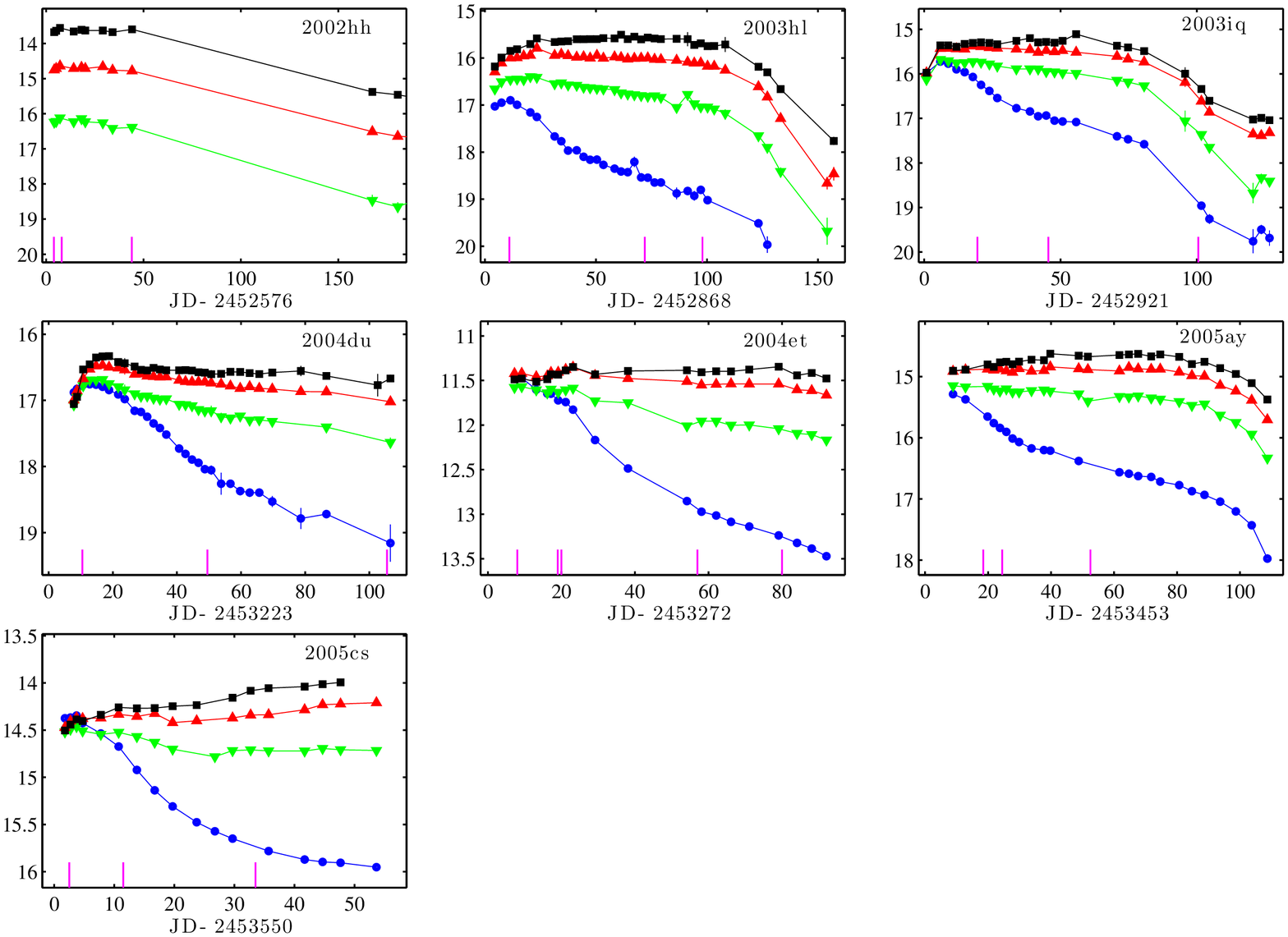}
\caption{Light curves of the 19 \snep\ from KAIT: $B$ (blue dots), $V$
(green upside-down triangles), $R$ (red triangles), and $I$ (black
squares). Vertical line segments at the bottom of each panel mark the
times for which we have spectra. Note the abundance of spectra, and
the diversity in the SN light-curve shapes.\label{f:LC}}
\end{figure}

Figure \ref{f:LC} presents the $BVRI$ light curves of our 19 SNe, and
shows the epochs at which spectra were obtained. In total, there are
about 1500 photometric points and 75 spectra used in this study. A
thorough analysis of these data is beyond the scope of this paper, and
will be discussed by Poznanski et al. (2009, in prep.). However, we
note the great diversity of plateau shapes, a subject to which we
return in \S\ref{s:culling}.

From the light curves we determine, using linear interpolation where
necessary, the apparent magnitude in the $I$ band and the \VmI\ color
at 50~d past explosion. We set the explosion date conservatively as
the midpoint between the last nondetection of the SN and its discovery
date (and the uncertainty as half this interval), except for
SN\,2005cs where we use the constraint determined by
\citet{pastorello06}. For many objects better constraints on the
explosion date could be set by using additional techniques (e.g.,
fitting templates to the photometry and/or spectroscopy); however, the
explosion-date uncertainties have little impact on our results, since
neither the color nor the magnitude (or \fe\ velocities, see below)
vary significantly during the plateau phase.  We have well-sampled
light curves, and as can be seen in Figure\ \ref{f:spec}, we have
spectra near day 50 for most of our SNe.

Since the KAIT filter system differs somewhat from the Bessell
curves used by N06, we also $S$-correct \citep{stritzinger02} 
the photometry to the standard Bessell 
\bvri curves taken from the database of \citet{moro00}. 
We do the correction by warping a standard day
50 template spectrum to the KAIT photometry, and measuring synthetic
magnitudes using the Bessell filter response. This has no noticeable
influence on our results, since the corrections are all smaller than
2\% in the measured values.

From the spectra we measure the apparent \fe\ velocity, defined here
as the minimum of the \fe\ line with respect to
5169\,\AA. Unfortunately, there is no ``industry standard'' framework
to determine the velocities of such broad, often noisy, and sometimes 
asymmetric lines; the different approaches we tried reveal systematic
differences in the derived values and hence different calibration
parameters. In an effort to develop a method that is simple, robust,
and readily available to the community, we decided to use the supernova
identification code \citep[SNID;][]{blondin07}, originally designed
for the automatic spectral classification of SNe. It uses the
\citet{tonry79} cross-correlation algorithm, and is flexible enough to
be used for our purpose.

\begin{figure}
\plotone{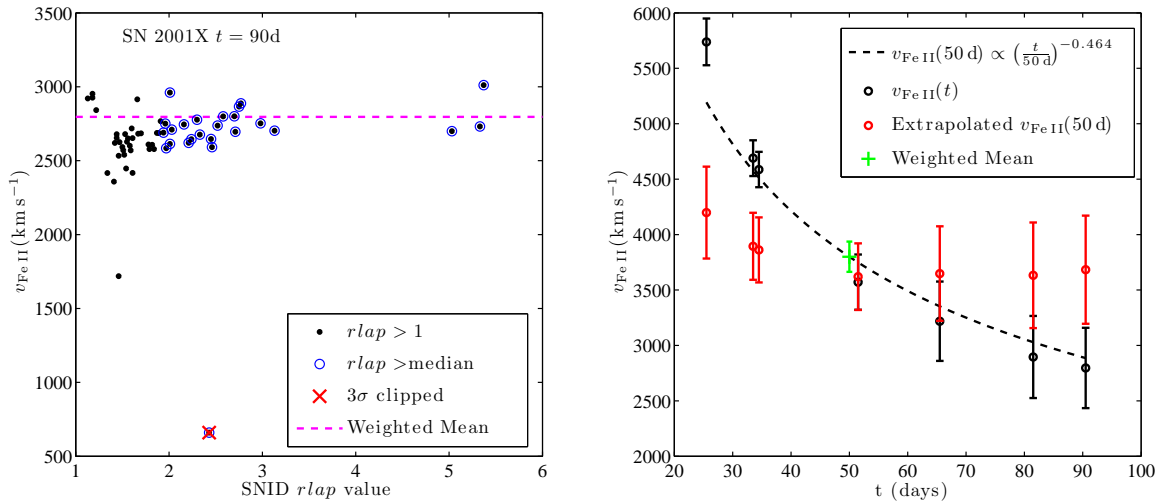}
\caption{
{\it Left:} Example application of our SNID-based \fe\ velocity
measurement to a typical spectrum.  A day 90 spectrum of SN\,2001X is
cross-correlated with a set of templates using SNID, and a velocity is
derived from each template (black dots).  The best-fitting templates
are selected based on SNID \textit{rlap} values (blue circles), and
sigma clipped to reject outliers (red crosses).  A weighted mean is
calculated from the selected velocities.  {\it Right:} For a given
object (SN\,2001X in this example), each velocity (black circles) is
propagated to day 50 (red circles) using Equation~(2) of N06 (dashed
line; our Equation (\ref{eq:nu2})).  A weighted mean supplies the final \fe\
velocity for that SN, and its uncertainty.  An additional uncertainty
of 150\,km\,s$^{-1}$ is added in quadrature to every SN, to account
for peculiar motion.\label{f:snid}}
\end{figure}

Figure \ref{f:snid} shows an example application of our method.  We use
the high-quality \snp\ templates distributed with SNID, for which the
\fe\ absorption-minimum wavelengths can be measured in a
straightforward way (e.g., by fitting a Gaussian to the line). We then
cross-correlate every spectrum with the templates (using only the
relevant wavelength range, 4500--5500\,\AA), choose only those
templates that produce a good fit (those with SNID $rlap$ values
greater than the median for the group), and calculate a sigma-clipped
($3\sigma$) weighted mean (left panel of Fig. \ref{f:snid}). We
convert the wavelength offsets to velocity using the relativistic
Doppler formula. Each velocity (and its uncertainty) is then propagated
to day 50 on the plateau using Equation~(2) of N06,
\begin{equation}
	v_{\mathrm{Fe\,II} }(50\,{\rm d}) = v_{\mathrm{Fe\,II} }(t)(t/50\,{\rm 
d})^{0.464\pm0.017}. \label{eq:nu2}
\end{equation}

Since most SNe have more than one available spectrum, we calculate for
each SN the weighted mean of all derived \fe\ velocities (right panel
of Fig. \ref{f:snid}). We add to the velocity uncertainty of every
SN a value of $\sigma_{\rm pec} = $150\,km\,s$^{-1}$, in quadrature,
to account for unknown peculiar velocities within the hosts
\citep{sofue01}. We have also confirmed that Equation (\ref{eq:nu2}) is
consistent with our spectra, and we find no substantial residual from
the N06 result, within its uncertainty.

As for the HPN sample, the apparent heliocentric redshifts of the SN
hosts are derived from NED\footnote{The NASA/IPAC Extragalactic
Database is operated by the Jet Propulsion Laboratory,
California Institute of Technology, under contract with the National
Aeronautics and Space Administration.}  and corrected for local flows
using the model of \citet{tonry00}. The properties of SNe in our sample,
as measured above, are included in Table \ref{t:KAIT}.

\subsection{Sample Comparison}\label{s:compsamp} 

\begin{figure}
\epsscale{0.5} \plotone{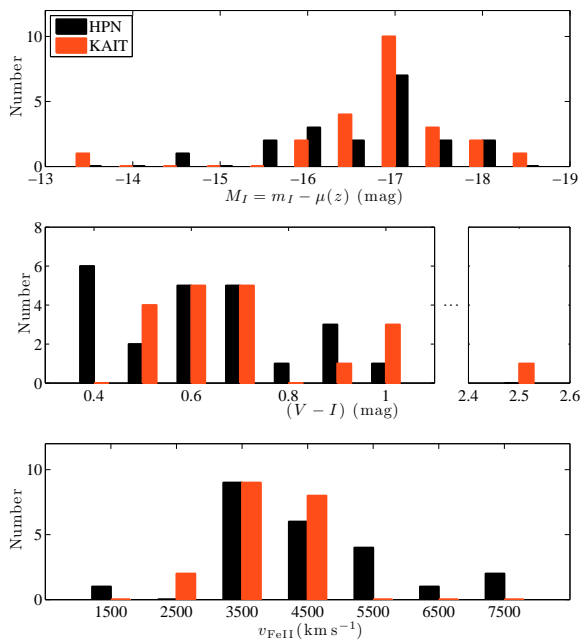} \caption{Comparison of the HPN and
KAIT samples, showing observables relevant for distance
measurements. Uncalibrated absolute magnitude 
(i.e., without applying dust or \fe\ velocity correction; top), \VmI\ (middle), and
\fe\ velocity (bottom).  KS statistics indicate that the samples can be
assumed to be drawn from the same underlying distribution at high
significance for the absolute magnitude, but quite low significance for the
color and \fe\ velocities.
\label{f:hist}}
\end{figure}

To understand whether HPN and KAIT sample a similar population, we
compare the relevant observables in both samples. Figure \ref{f:hist}
shows a comparison of the distribution of uncorrected absolute
$I$-band magnitudes (top panel), $V-I$ colors (middle), and \fe\
velocities (bottom), all at day 50 on the plateau, as measured
above. These are the quantities used for standardization and distance
measurement. Surprisingly, the luminosity distributions appear consistent with
each other, while the KAIT sample features redder colors indicating a higher
average extinction, and lower velocities that point to lower
luminosities.

Using Kolmogorov-Smirnov (KS) statistics, we find that the luminosity
distributions match reasonably well ($P_{\rm K-S}$ values of 0.66),
whereas the colors and velocities are less consistent ($P_{\rm K-S}$
values of 0.13 and 0.06, respectively). Those results remain similar
if one considers only $z<0.1$ SNe.  However, the high-velocity objects
in the HPN sample all have substantial uncertainties that are not
taken into account by KS statistics. For example, most of the 
highest-velocity SNe in the HPN samples have velocity uncertainties
of 2000\,km\,s$^{-1}$. Nevertheless, differences in
sample properties should be expected, due to the era in which they
were compiled. KAIT is expected to be sensitive to fainter SNe, either
intrinsically (hence with lower average velocities) or due to dust
obscuration (hence with red colors).  One should note that while the
KAIT sample may be less biased than the HPN sample, the luminosity
function is still skewed significantly toward bright objects, due to
follow-up criteria, as the median absolute magnitude is 0.5$-$1 mag
brighter than the observed distribution of all KAIT \snep\ (i.e.,
including those SNe that were not scheduled for follow-up
observations; Li et al. 2009, in prep.).  While the importance of this
bias should be explored with future more balanced samples, we note
that at high redshift the same selection naturally occurs. The
difference between the samples in terms of the observed properties has
a visible effect on the Hubble diagram we construct, as discussed in
\S\ref{s:bestfit}.

With an absolute magnitude distribution spanning more than 5 mag and a
diverse set of light-curve shapes (Fig.\ \ref{f:LC}), the
standardizability of \snep\ would appear to be a significant
challenge. However, as we demonstrate below, the resultant Hubble
diagram remains tight, and the velocity vs. luminosity correlation
persists, despite the larger and less homogeneous sample.

Using the 23 SNe in the HPN sample, and the 19 in the KAIT sample, we
construct our list of 40 SNe, after merging the two shared SNe,
SN\,1999em and SN\,1999gi. For these two SNe we use the KAIT
measurements, due to the somewhat smaller uncertainties (but
consistent values) we obtain for the observables. Table \ref{t:final}
lists the full sample used in the subsequent analysis.

\begin{deluxetable}
{lrrrrrl} 
\tabletypesize{\scriptsize}


\tablecaption{Final Sample \label{t:final}} 
\tablewidth{0pt} 
\tablehead{ 
\colhead{Name\tablenotemark{a}}  &
\colhead{$cz/z$\tablenotemark{b}} &
\colhead{$v_{\mathrm{Fe\,II} }$\tablenotemark{c}} &
\colhead{$m_I$\tablenotemark{c}} &
\colhead{$V-I$\tablenotemark{c}} &
\colhead{Best-fit $\mu$} &
\colhead{Source} \\
\colhead{} &
\colhead{(km\,s$^{-1}$)} &
\colhead{(km\,s$^{-1}$)} &
\colhead{(mag)} &
\colhead{(mag)} &
\colhead{(mag)} } 
 \startdata 
SN2002hh &   297(187) &  3692( 435) & 13.68(0.05) & 2.67(0.05)  & 28.89(0.23) & KAIT \\ %
SN2004et &   297(187) &  3980( 194) & 11.39(0.01) & 0.52(0.01)  & 28.36(0.09) & KAIT \\ %
SN2005cs &   505(187) &  2102( 211) & 13.98(0.02) & 0.69(0.03)  & 29.61(0.21) & KAIT \\ %
SN1999gi &   643(187) &  3697( 200) & 14.00(0.02) & 0.86(0.03)  & 30.57(0.13) & KAIT \\ %
SN1999br &   757(187) &  1545( 300) & 16.67(0.05) & 0.84(0.07)  & 31.60(0.43) & N06 \\ %
SN2005ay &   816(187) &  3432( 224) & 14.67(0.02) & 0.63(0.06)  & 31.27(0.13) & KAIT \\ %
SN1999em &   819( 70)\tablenotemark{d} &  3382( 171) & 13.23(0.03) & 0.58(0.03)  & 29.84(0.11) & KAIT \\ %
SN1991G  &  1029(187) &  3347( 500) & 15.01(0.09) & 0.45(0.11)  & 31.70(0.36) & N06 \\ %
SN1992ba &  1064(187) &  3523( 300) & 14.65(0.05) & 0.59(0.07)  & 31.33(0.18) & N06 \\ %
SN1989L  &  1189(187) &  3529( 300) & 14.47(0.05) & 0.88(0.07)  & 30.94(0.19) & N06 \\ %
SN1986I  &  1190(187) &  3623( 300) & 13.98(0.09) & 0.45(0.22)  & 30.82(0.29) & N06 \\ %
SN1990E  &  1273(187) &  5324( 300) & 14.51(0.20) & 1.31(0.28)  & 31.44(0.37) & N06 \\ %
SN2001x  &  1329(187) &  3800( 203) & 14.74(0.02) & 0.57(0.03)  & 31.59(0.11) & KAIT \\ %
SN1990K  &  1623(187) &  6142(2000) & 13.87(0.05) & 0.58(0.21)  & 31.62(0.75) & N06 \\ %
SN2003hl &  2105(187) &  3561( 293) & 15.60(0.02) & 0.98(0.04)  & 32.01(0.14) & KAIT \\ %
SN2003iq &  2105(187) &  4043( 179) & 15.27(0.03) & 0.66(0.03)  & 32.16(0.10) & KAIT \\ %
SN2001bq &  2238(187) &  4023( 238) & 15.55(0.03) & 0.51(0.08)  & 32.55(0.13) & KAIT \\ %
SN1999ca &  2772(187) &  5353(2000) & 15.56(0.05) & 0.73(0.07)  & 32.93(0.89) & N06 \\ %
SN2001cm & 0.011(0.001) &  4065( 189) & 16.30(0.02) & 0.68(0.03)  & 33.18(0.10) & KAIT \\ %
SN1991al & 0.013(0.001) &  7330(2000) & 16.06(0.05) & 0.39(0.07)  & 34.28(0.54) & N06 \\ %
SN2000dj & 0.015(0.001) &  4784( 250) & 17.00(0.09) & 0.59(0.11)  & 34.26(0.18) & KAIT \\ %
SN1992af & 0.016(0.001) &  5322(2000) & 16.46(0.20) & 0.43(0.28)  & 34.05(0.87) & N06 \\ %
SN2004dh & 0.017(0.001) &  4990( 300) & 17.60(0.15)\tablenotemark{e} & 0.70(0.21)\tablenotemark{e}  & 34.86(0.30) & N06 \\ %
SN1999cr & 0.019(0.001) &  4389( 300) & 17.44(0.05) & 0.56(0.07)  & 34.56(0.18) & N06 \\ %
SN1999eg & 0.019(0.001) &  4012( 300) & 17.72(0.05) & 0.55(0.07)  & 34.68(0.16) & N06 \\ %
SN1993A  & 0.027(0.001) &  4290( 300) & 18.56(0.05) & 0.51(0.07)  & 35.68(0.17) & N06 \\ %
SN2000bs & 0.028(0.001) &  4115( 276) & 17.67(0.17) & 0.51(0.20)  & 34.71(0.31) & KAIT \\ %
SN1993S  & 0.029(0.001) &  4569( 300) & 18.22(0.05) & 0.69(0.07)  & 35.32(0.14) & N06 \\ %
SN1992am & 0.042(0.001) &  7868( 300) & 17.90(0.05) & 0.38(0.07)  & 36.27(0.11) & N06 \\ %
SN04D4fu & 0.119(0.001) &  3861( 150) & 21.99(0.04) & 0.59(0.05)  & 38.85(0.11) & N06 \\ %
SN03D4cw & 0.138(0.001) &  3067( 660) & 22.33(0.09) & 0.90(0.10)  & 38.52(0.47) & N06 \\ %
SN04D1pj & 0.139(0.001) &  4981( 214) & 21.99(0.04) & 0.66(0.06)  & 39.28(0.11) & N06 \\ %
SN04D1ln & 0.185(0.001) &  3593( 159) & 22.79(0.05) & 0.70(0.07)  & 39.43(0.12) & N06 \\ %
SN03D3ce & 0.257(0.001) &  5762( 522) & 23.45(0.50) & 0.39(0.50)  & 41.22(0.79) & N06 \\ %
\\
\underline{Culled SNe}\\
\\
SN1999bg &  1214(187) &  4657( 439) & 15.31(0.02) & 0.74(0.04)  & 32.41(0.18) & KAIT \\ %
SN2002gd &  2418(187) &  2774( 315) & 16.78(0.03) & 0.74(0.05)  & 32.90(0.21) & KAIT \\ %
SN2000dc & 0.010(0.001) &  4545( 232) & 16.05(0.06) & 0.97(0.06)  & 32.93(0.14) & KAIT \\ %
SN2001do & 0.010(0.001) &  3927( 311) & 15.75(0.03) & 0.97(0.05)  & 32.35(0.15) & KAIT \\ %
SN2001cy & 0.015(0.001) &  3987( 218) & 16.13(0.04) & 0.63(0.05)  & 33.01(0.12) & KAIT \\ %
SN2004du & 0.017(0.001) &  4855( 302) & 16.60(0.04) & 0.53(0.07)  & 33.94(0.13) & KAIT \\ %
\enddata 
\tablenotetext{a}{Sorted by increasing redshift.}
\tablenotetext{b}{From local Hubble-flow model of \citet{tonry00} (unless 
otherwise noted). For $cz>3000$ km\,s$^{-1}$ the redshift is given.} 
\tablenotetext{c}{At day 50, on the plateau.} 
\tablenotetext{d}{From the Cepheid distance of \citet{leonard03} and the Hubble law.} 
\tablenotetext{e}{Note that these numbers differ from those in N06, 
following our re-reduction of the data.} 
\vspace{0.5cm}
\end{deluxetable}

\section{Fitting Method}\label{s:method}

N06 assume a correlation between luminosity and Fe~II velocity that
is only modified by a \citet{cardelli89} extinction law, with
$R_V=3.1$. However, there is no {\it a priori} reason to assume that
all of the SNe in the sample are subject to the same extinction law, with
the average curvature found in the Milky Way.  For that reason we
modify the color-dependent term in Equation (1) of N06 to be a free
parameter. This term, which has a value of 1.36 in N06\footnote{Note
the sign misprint in N06 for that term.}, we refer to as $R_I$. Given the
color excess $E(V-I)\equiv(V-I)-(V-I)_0$, $R_I$ defines the amount of
extinction in the $I$ band, such that $A_I = R_I E(V-I)$. The value of
$R_I$ can be directly derived from the \citet{cardelli89} extinction
curve as a function of $R_V$. We therefore obtain a slightly modified
Equation (1) of N06:
\begin{equation}
\label{eq:M} M_I= M_{I_0}-\alpha \log_{10}(v_{\mathrm{Fe\,II} }/5000) +R_I \left[ (V-I)-(V-I)_0 \right]. 
\end{equation}

In addition, we use ``Hubble-constant free'' absolute magnitudes and
luminosity distances ${\cal M}_{I_0}\equiv M_{I_0}-5\log_{10}
(H_0)+25$ and ${\cal D}_{L} \equiv H_0 D_L$
\citep{goobar95,sullivan06a}, since we use no anchor of ``absolute''
distance, and measure only the shape of the Hubble diagram (relative
distances). We thus find
\begin{equation}
\label{eq:dl} {\cal M}_{I_0}-\alpha \log_{10}(v_{\mathrm{Fe\,II} }/5000) +R_I \left[ (V-I)-0.53 \right] -m_I = -5\log_{10}({\cal D}_L(z | \Omega_m ,\Omega_\Lambda)), 
\end{equation}
where we use for the ridge-line color the value from N06, $(V-I)_0 = 0.53$
mag. As mentioned by N06, the precise value for this color is degenerate 
with ${\cal M}_{I_0}$, and does not influence our other results. As seen in the middle panel of 
fig. \ref{f:hist}, perhaps due to the ubiquity of dust around \snep, or as a consequence of an intrinsic dispersion, 
it is difficult to find a precise value for the unreddened color of those SNe. 

Since all of the terms in Equation (\ref{eq:dl}) have errors,
including significant redshift uncertainties for the nearby SNe, we
maximize the marginal likelihood ${\cal L}$ by minimizing the
multivariate cost function

\begin{equation}
\begin{split}
-2\log ({\cal L})&=\sum\{\, \frac{[{\cal M}_{I_0}-\alpha \log_{10}(v_{\mathrm{Fe\,II} }/5000) +R_I \left[ (V-I)-0.53 \right] -m_I + 5\log_{10}({\cal D}_L(z))]^2} { \alpha \sigma^2_{\log_{10}(v_{\mathrm{Fe\,II} }/5000)} +R_I \sigma^2_{(V-I)} +\sigma^2_{m_I} + \sigma^2_{5\log_{10}({\cal D}_L(z))} + \sigma^2_{\rm sys}} \\
& + \log(\alpha \sigma^2_{\log_{10}(v_{\mathrm{Fe\,II} }/5000)} +R_I \sigma^2_{(V-I)} +\sigma^2_{m_I} + \sigma^2_{5\log_{10}({\cal D}_L(z))}+ \sigma^2_{\rm sys} ) \,\}, \label{eq:ll} 
\end{split}
\end{equation}
where the summation is over the SNe in the sample \citep[e.g.,
][]{kelly07}. The second, logarithmic term --- which comes from the
normalization of the likelihood function --- helps mitigate the
tendency of the first term to overfit the data by preferring large
values of $\alpha$, $R_I$, and $\sigma_{\rm sys}$.  We note, however,
that this term does not significantly change the resulting best-fit
parameters. We include in the denominator (and the logarithmic term)
of Equation (\ref{eq:ll}) a systematic uncertainty $\sigma_{\rm sys}$
in order to include (and measure) the intrinsic scatter in the
correlation (i.e., the contribution to the dispersion not accounted
for by the measurement errors).

The parameters to be fit are therefore ${\cal M}_{I_0}$ (although for
simplicity we will keep citing values for $M_{I_0}$, assuming
$H_0=70$\,km\,s$^{-1}$\,Mpc$^{-1}$), $\alpha$, $R_I$ (that we translate
to $R_V$), $\sigma_{\rm sys}$, and the cosmological parameters. For
the present low-$z$ data, since we do not have any leverage on the
cosmological parameters, we assume a standard cosmology
($\Omega_m=0.3$, $\Omega_\Lambda=0.7$), and only fit for $\sigma_{\rm
sys}$ and the correlation coefficients.

\pagebreak
\section{Results}\label{s:results}

Using the complete sample of 40 SNe, we find that the best-fit values
after marginalization are $\alpha=4.6\pm0.7$, $R_V=1.5\pm0.5$
($R_I=0.7^{+0.3}_{-0.4}$), $M_{I_0}=-17.43\pm0.10$\,mag, and
$\sigma_{\rm sys}=0.38$\,mag.  The systematic scatter is equivalent to
an 18\% uncertainty in distance, largely dominating the error budget,
being more than twice the median measurement uncertainty.  However,
about a third of that scatter is due to two SNe in the KAIT sample,
SN\,2001cy and SN\,2001do, and to some lesser extent SN\,1999bg.
Those SNe are offset from the best-fit solution by $6.7\sigma$,
$4.1\sigma$, and $3.6\sigma$, respectively.  An analysis excluding
those objects gives a tight correlation with $\sigma_{\rm
sys}=0.22$\,mag, which corresponds to 10\% in distance.

\subsection{Sample Culling for Cosmology} \label{s:culling}

\begin{figure}
\epsscale{1} \plotone{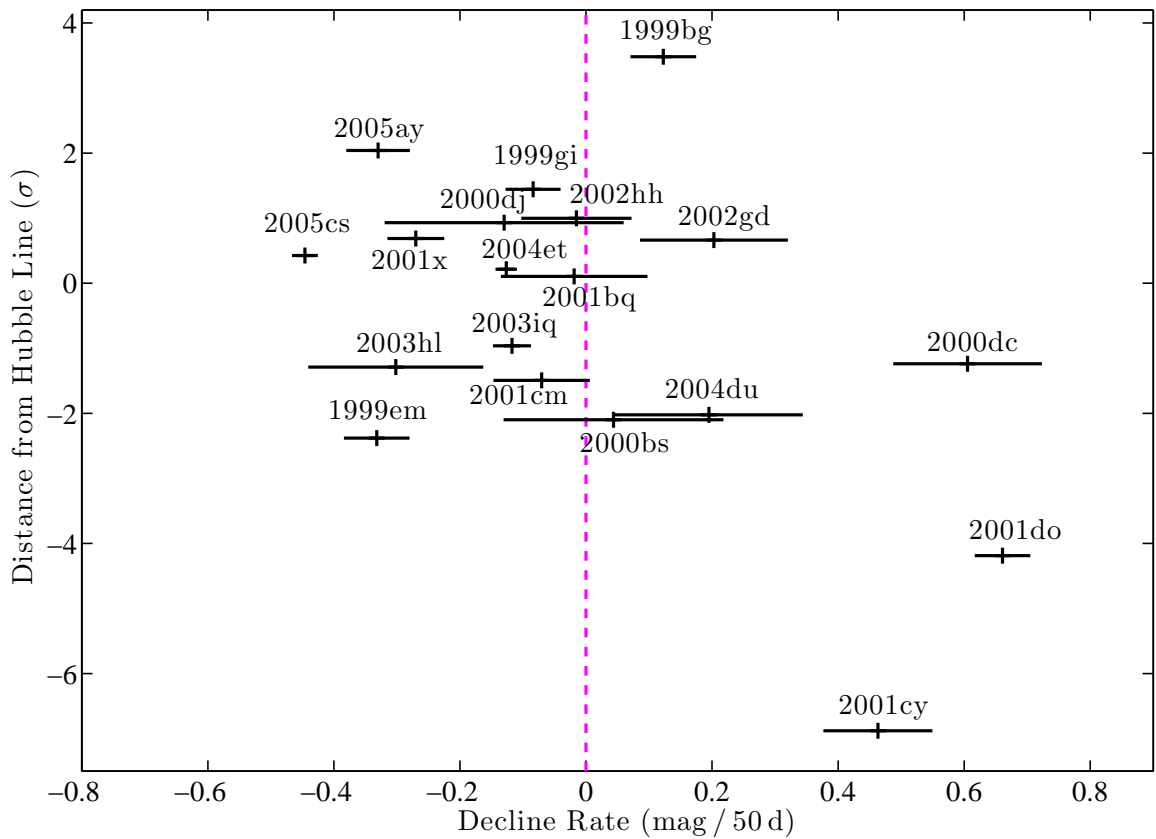}
\caption{Offset from the Hubble law fit to the full sample before
culling, as a function of the $I$-band decline rate.  As can be
clearly seen, SN\,2001cy and SN\,2001do are amongst the strongest
decliners and are our most significant outliers, followed by
SN\,1999bg which declines as well. The magenta dashed line marks our
cutoff, where we select only SNe that have decline rates smaller than
(or consistent to 1$\sigma$ with) zero.
\label{f:decline}}
\end{figure}

A possible glimpse at the source of the discrepancy lies in the
photometric behavior of the outlying SNe. This is portrayed in Figure
\ref{f:decline}, where we show, for the KAIT SNe, the distance from
the Hubble law (as derived from the complete sample) in $\sigma$
units, as a function of the decline rate in the $I$ band.  We define
here the decline rate as the slope of a linear fit to the $I$-band
light curve starting 10~d after explosion (after the rise time), and
ending at day 50, in units of mag (50~d)$^{-1}$. The errors are
estimated using a Monte-Carlo simulation propagating the uncertainties
in the photometry and in the explosion dates.

Figure \ref{f:decline} shows that SN\,2001cy and SN\,2001do have
substantial decline rates, about 0.5~mag (50~d)$^{-1}$, while most
objects barely decline, or even have rising light curves. The decline
vs. offset correlation is not strong; consequently, adding a decline
parameter to the fitting only marginally improves the scatter to
$\sigma_{\rm sys}=0.33$\,mag.  However, we can now define
quantitatively a selective subsample of \snep\ for the
purpose of distance measurement in this paper, and reject objects in a
controlled manner.  For the purpose of this work, we will keep objects
that have decline rates smaller than (or consistent to 1$\sigma$ with)
zero.\footnote{Note that we are defining the specific subsample of \snep\ 
in this way only for the purpose of having a clear and clean set for
cosmological purposes. 
We do {\it not} mean to imply that every
SN\,II having a decline rate larger than zero is necessarily a
SN\,II-L or some other ``not-II-P'' SN. 
The real distinction between 
\snep, SNe\,II-L, and other potential SN\,II subtypes will
be considered in more detail by Poznanski et al. (2009, in prep.).}

As can be seen in Figure \ref{f:decline}, this criterion rejects all
three outliers (SNe\,2001cy, 2001do, and 1999bg) by design, but also
SNe\,2000dc, 2002gd, and 2004du that are relatively well fit.
This choice leaves 34 SNe in the combined sample that we analyze in
the following sections.  We emphasize that a similar cut has been
applied by N06 prior to the costly follow-up observations of the
moderate-$z$ SNe in their sample (P. Nugent, 2008, private comm.), and
has most probably been applied to the HP02 sample as well. The precise
criteria have not been detailed previously, and could significantly
bias the best-fit parameters and dispersion. Our choice, while
somewhat arbitrary, does not affect our results below, and any choice
that rejects the two major outliers will effectively be equivalent.

\subsection{Best-Fit Parameters and Hubble Diagram}\label{s:bestfit}

\begin{figure}
\epsscale{.75} \plotone{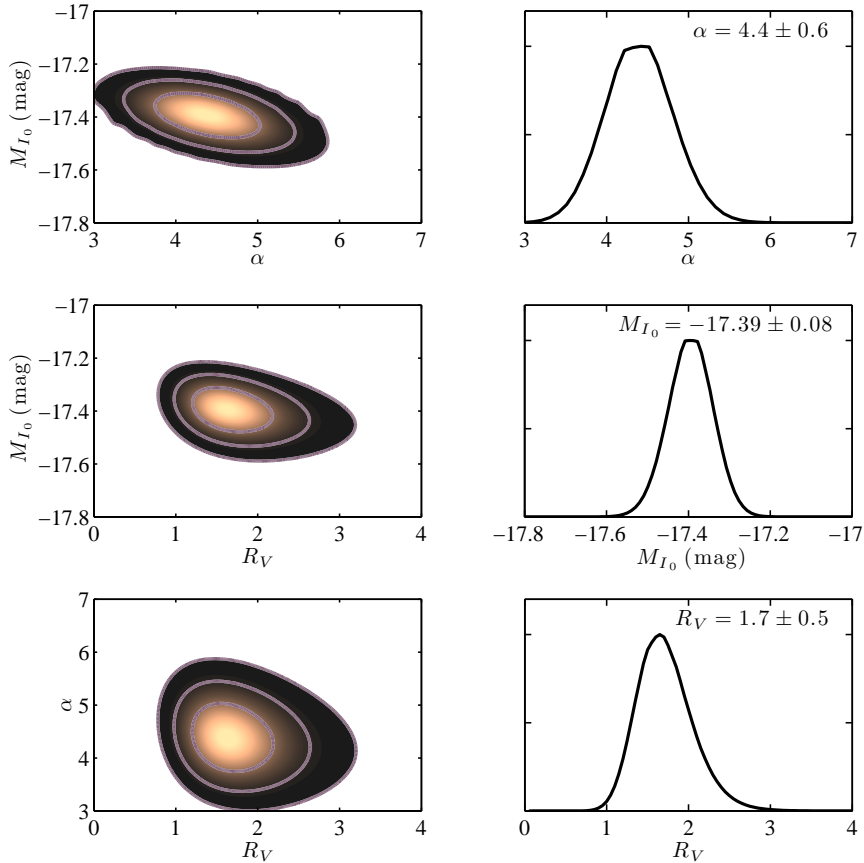} 
\caption{Best-fit parameter ($\alpha,M_{I_0},R_V$) $1\sigma$,
$2\sigma$, and $3\sigma$ contours at left; marginalized posterior
probabilities at right, for the culled sample of 34 SNe.
\label{f:contours}}
\end{figure}

Using the culled sample of 34 SNe, we find that the best-fit values
after marginalization are $\alpha=4.4\pm0.6$, $R_V=1.7\pm0.5$
($R_I=0.8\pm0.3$), $M_{I_0}=-17.39\pm0.08$\,mag, and $\sigma_{\rm
sys}=0.22$\,mag (10\% in distance), consistent with the values for the
complete sample.  The likelihood contours and the marginalized
posteriors are plotted in Figure \ref{f:contours}. As can be seen, the
coefficients are weakly covariant.

The value of the best-fit average $R_V$, while small, is consistent
with results for many SNe Ia, which tend to suffer an extinction that
is very selective
\citep{elias-rosa06,krisciunas07,elias-rosa08,nobili08,wang08}. As
shown by \citet{wang05} and \citet{goobar08}, this result could be
explained with a fairly simple model where ``normal,'' yet
circumstellar, dust causes heavily color-dependent scattering. While
this model is discussed in the context of SNe\,Ia, core-collapse SNe
are expected (and observed) to be enshrouded by more circumstellar
dust, which could explain these low $R_V$ values. We note that a small
$R_V$ is still the best fit if one removes the highly extinguished
SN\,2002hh from the analysis.

\begin{figure}
\epsscale{.8} \plotone{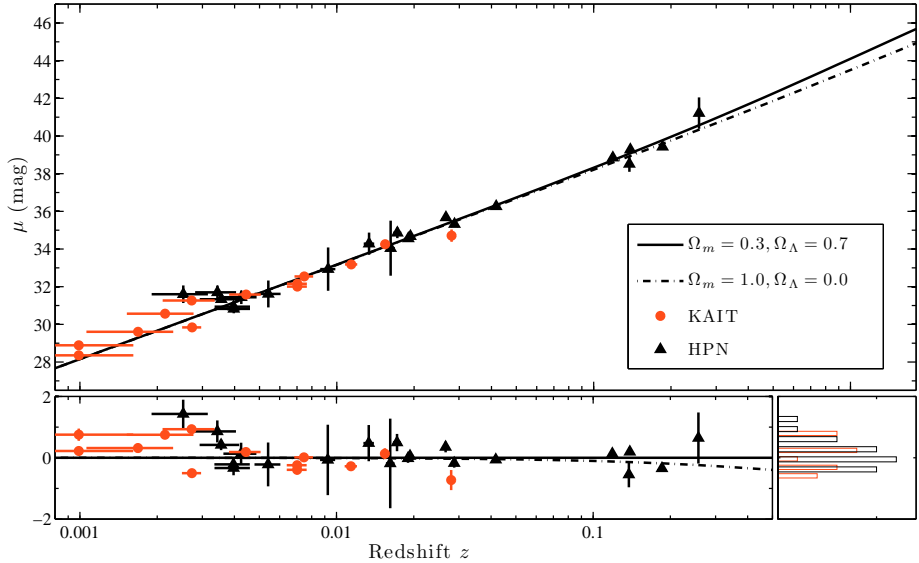} 
\caption{Hubble diagram (top), and residuals from standard concordance
cosmology (bottom), for the HPN sample (black triangles) and the culled KAIT
sample (orange filled circles). ``Standard'' (line) and
Einstein-deSitter (dash-dot) cosmologies are added to guide the
eye. The bottom-right panel shows the similarity in the distribution
of residuals for both samples.\label{f:HD}}
\end{figure}

Using Equation (\ref{eq:dl}) we build the Hubble diagram in Figure
\ref{f:HD}.
The scatter around the standard cosmology trend is quite small, and is
accounted for by the systematic and measurement uncertainties.  The
systematic scatter can be contained in a reasonable velocity
dispersion of $\sigma_{v_{\mathrm{Fe\,II} }}=135$\,km\,s$^{-1}$. This
is equivalent to, or better than, the results obtained by previous
authors, and is encouraging considering the size of the sample and its
initial inhomogeneity.

\subsection{Robustness of the Correlation}\label{s:robust}

In order to test the robustness of the correlation, we explore the
importance of the different ingredients in the fit. If one applies
{\it no correction at all} (i.e., assume the \snep\ are perfect
standard candles), the correlation crumbles, and gives $\sigma_{\rm
sys} \approx 0.9$ mag. The same occurs if one does not apply the
velocity correction, but only the color correction. 
A simple test of the importance of that term is
obtained by shuffling the velocity measurements among the objects,
assigning to each SN a random value chosen from the sample
distribution. The result, on average, is a very weak correlation, with
$\sigma_{\rm sys} \approx 0.6$ mag, and unrealistic best values for
the coefficients ($\alpha<0$ and $R_V>15$ in most cases; see \citealt{draine03} for a review on 
measured values and theoretical limits on $R_V$).

However, if we use solely the \fe\ correction 
(which is numerically equivalent to setting $R_I=0$ or $R_V=0.78$),
we get a scatter only slightly greater than for the best solution
($\sigma_{\rm sys} \approx 0.26$ mag), and mildly different values for
$\alpha$ and $M_{I_0}$. This result indicates that for our
sample of SNe, the dust correction is not strongly required by
the data.

Since we {\it do} expect SN magnitudes to be affected by dust,
there are a few possible explanations we consider. First, it could be
that an intrinsic color-velocity correlation masks most of the
contribution from dust. While theoretically one could expect the color
(i.e., the temperature) to be correlated with the photospheric
velocity at least to some extent, we find no indication in the sample
for any such covariance, despite having at least a few SNe suffering
negligible extinctions.

An alternative explanation is that the sample is heavily biased
toward dust-free objects. While this is securely wrong for at least
one object \citep[SN\,2002hh, which suffers $\sim$5 mag of
extinction in the $V$ band;][]{pozzo06}, it is probably wrong for many
of the other SNe as well. H05 finds significant dust corrections for at
least some of the SNe in his sample, and our sample should be
less ``hand-picked'' as the objects included were discovered by modern-era
CCD-equipped SN searches that can find more \snep\ buried in their
host galaxies.

However, neglecting SN\,2002hh, most SNe in the sample are within
$\sim0.3$\,mag of the sample's mean \VmI\ color. Additionally, for any
reasonable dust law, the color term in Equation (\ref{eq:dl}) is at
least 3--5 times less significant than the velocity term, so that even
substantial differences in color will contribute relatively little to
the dispersion in distance moduli. Consequently, a sample that does
not have many heavily extinguished objects can be fit about as well
when assuming no extinction at all. 

We have searched unsuccessfully for parameters that correlate with the
residuals from the Hubble diagram and further reduce the necessary
$\sigma_{\rm sys}$ value, other than the slope of the plateau in the
light curve that we have used to cull our sample in \S\ref{s:culling}.
The $B$-band luminosity at day 50, for example, does not reduce the
scatter by more than a few percent of its previous value.  There are
indications that rejection of red objects (those having $(V-I)-(V-I)_0
\ga 0.2$ mag) may reduce the scatter, but our sample is still too
small for us to make a robust statement regarding this.

We also determine the best-fit parameters for different subsamples of
the data. The HPN and KAIT samples when analyzed separately give
best-fit parameters consistent to within $1\sigma$ with those derived
from the full set. When examining various cuts in redshift, we find
that the lowest-redshift objects in the sample ($z \leq 0.004$) tend
to pull the solutions to somewhat larger values of $\alpha$ (near 6
instead of 5), as previously noted by N06 and tentatively ascribed to
a Malmquist bias. At $z \ga 0.004$ there seems to be a systematic difference
between the two samples, with the KAIT SNe being mostly underneath the 
Hubble-law line. The number of SNe in this range is too small for a 
conclusive analysis, but we estimate that this is a reflection 
of the differences between the samples noted in \S\ref{s:compsamp}. 
KAIT finds intrinsically fainter and more extinguished SNe, 
for which the fit compensates by reducing their derived distances.

\subsection{Shared-Host SNe} 

SN\,2002hh and SN\,2004et both occurred in the same host galaxy,
NGC\,6946, a neighbor of the Milky Way ($<10$\,Mpc away). While
SN\,2004et shows no apparent uniqueness (except perhaps some small
photometric jitter in the plateau phase of the light curve),
SN\,2002hh is highly reddened by dust, with $A_V \approx 5$ mag and an
infrared echo from a shell with $\sim$10 M$_{\odot}$
\citep{barlow05,pozzo06}.  Despite the extreme extinction, the
distances of the two SNe, as derived using the best-fit parameters of
\S\ref{s:bestfit}, are consistent within about $2\sigma$ (see Table
\ref{t:final}).

\begin{figure}
\epsscale{1} \plottwo{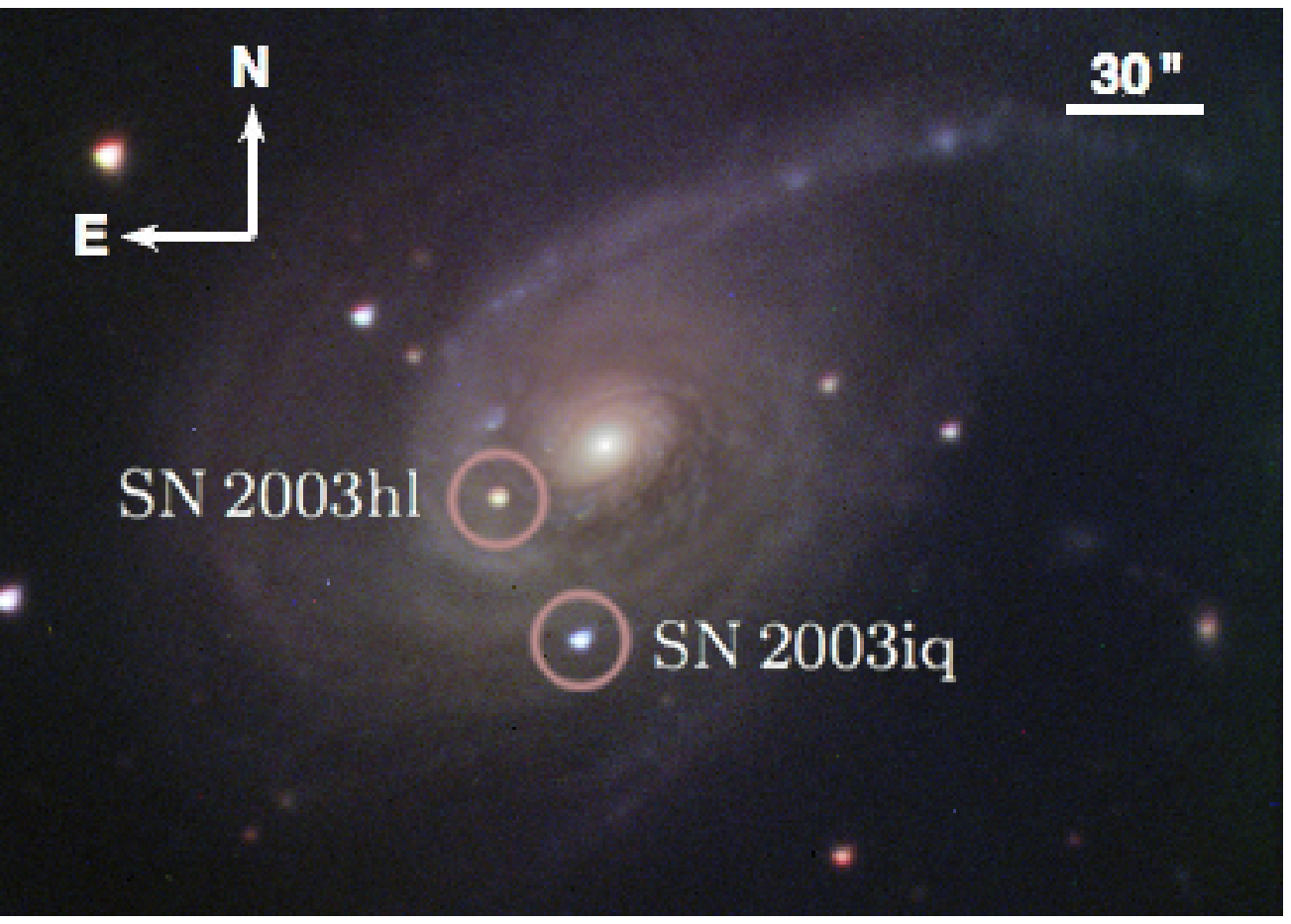}{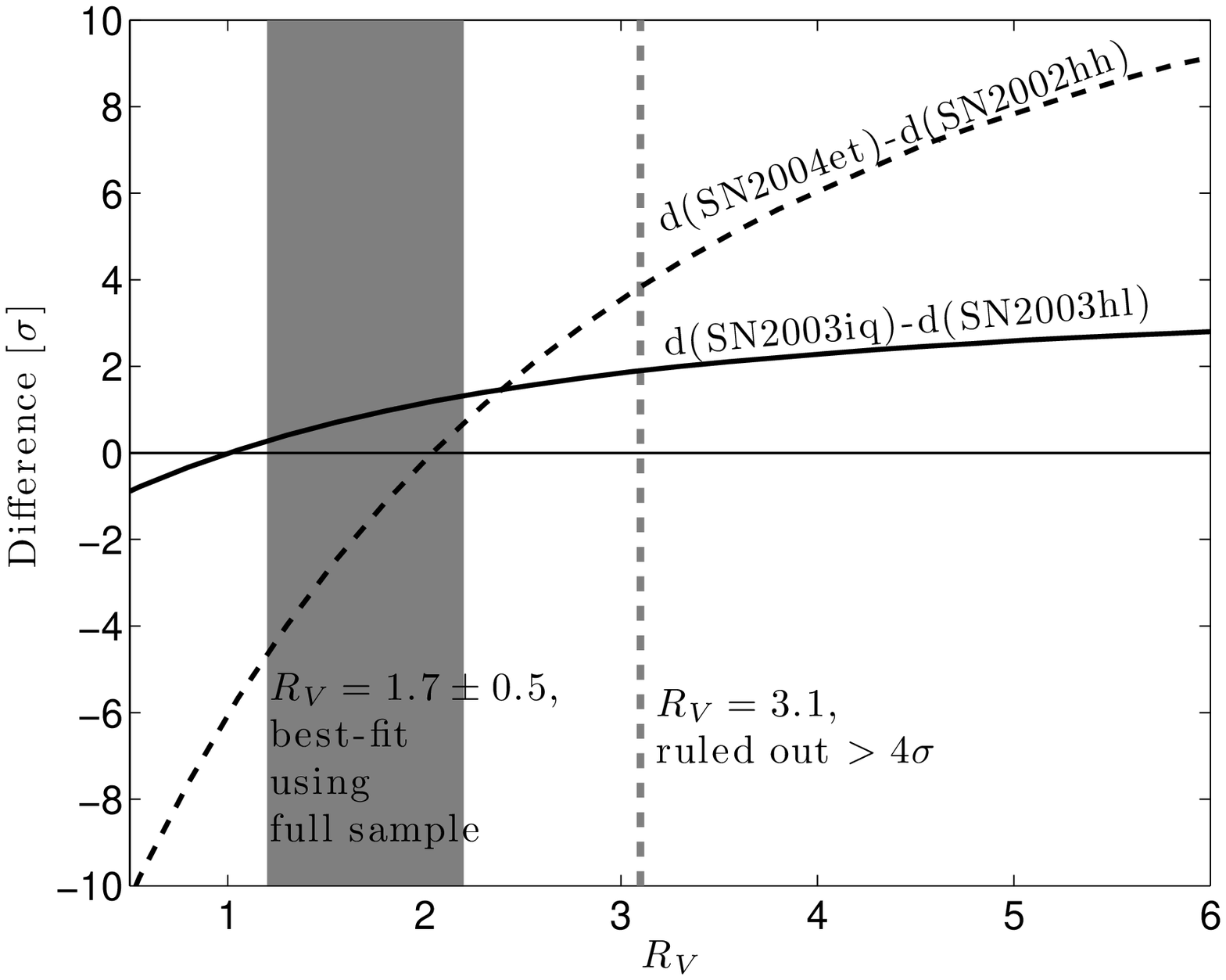}
\caption{
{\it Left:} Color composite image of NGC\,772 showing SN\,2003hl
and SN\,2003iq. It consists of KAIT \bvri images, and a
deeper frame from Deep-Sky (Nugent et al. 2009, in prep.) not showing
the SNe, but enhancing galaxy features.  {\it Right:} Distance difference
between SNe 2003hl and 2003iq (solid line) and between SNe 2002hh and
2004et (dashed line), in $\sigma$ units, as a function of the assumed
$R_V$.  A value of $R_V = 3.1$ (dashed grey line) is rejected at more
than $4\sigma$.  The gray area marks the $1\sigma$ interval preferred
by the full sample. These two pairs of SNe, as well as the full
sample, favor a low-$R_V$ dust model.
\label{f:ngc0772}} 
\end{figure}

SN\,2003hl and SN\,2003iq exploded within weeks of each other, in the
same host galaxy, NGC\,772 (see Fig. \ref{f:ngc0772}, left
panel). In fact, SN\,2003iq was discovered by an amateur SN observer while
following 2003hl \citep{llapasset03}. With the exception of the first
spectrum of SN\,2003hl, our spectra were obtained by placing the slit
on both objects simultaneously.  In this lower-extinction case
(compared with SN\,2002hh in NGC\,6946), the
agreement in distance is even better.

However, the distances to both pairs of SNe best agree only for the
favored model with low $R_V$. (A similar value for SN\,2002hh has been
measured by \citealt{pozzo06} using a full light-curve comparison to
SN\,1999em.)  As seen in the right-hand panel of Figure
\ref{f:ngc0772}, a value of $R_V=3.1$ is rejected at a combined
significance level higher than $4\sigma$.  This result strongly
supports the low $R_V$ value favored by the full sample.

\subsection{Error Budget: Prospects for Future Samples}

While the scatter we find in \S\ref{s:bestfit} is small, it is tightly
coupled to our error estimates. The value $\sigma_{\rm sys}=0.22$\,mag
is in effect the scatter unaccounted for by other error terms.  As can
be seen in Table \ref{t:final}, due to the quality of our data, our
uncertainties are generally smaller than those of previous authors. If
one argues that those uncertainties are underestimated (though we find
that they are not), increasing them by some amount will result in an
even smaller value of $\sigma_{\rm sys}$.

As a consequence of our precise measurements, systematic errors
dominate our Hubble diagram.  This allows us to constrain the
intrinsic dispersion of the correlation to be roughly 10\% in
distance. However, one should note that there are observational
limitations that would increase that uncertainty for any foreseeable
sample.  Even for data with exquisite photometry, and high S/N
spectroscopy, there still remains an uncertainty in the peculiar
velocity of the exploding star, on the order of
150\,km\,s$^{-1}$. This translates to about 0.05$-$0.08\,mag of
uncertainty, i.e., 2$-$4\% in distance. A reasonable precision to
expect for a SN with a single spectrum is about 300\,km\,s$^{-1}$, or
$\sim$0.12\,mag.

In addition, any project that wishes to amass a suitable sample of
\snep\ will need to address the critical issue of sample definition
discussed in \S\ref{s:culling}.  A simple strategy to overcome this
would include high-cadence observations during a ``detection phase,'' in
order to have strong constraints on the explosion dates, followed by
low-cadence monitoring of the candidates, in order to ascertain that
they do not decline.

The weak dependence on \VmI\ color discussed in \S\ref{s:robust} has
strong utility for extension to high-$z$ SNe, where selection against
faint and reddened SNe is stronger, and dust could have less impact.
A Hubble diagram based only on the $V$ band, without any color term,
is equivalent to fixing $R_I=-1$ in Equation (\ref{eq:dl}).  Solving for
the other parameters, we find a somewhat inferior scatter of
$\sigma_{\rm sys} = 0.29$ mag. However, obtaining rest-frame $I$-band
magnitudes for $z>0.4$ SNe is observationally challenging, requiring
near-infrared observations. The slightly poorer scatter in the $V$ band can
be overcome by statistics.  We do not expect significant scatter (or
evolution) in those photometric bands due to metallicity.  Figure 13
of \citet{baron03} shows that wildly different metallicities produce
little variance in the intrinsic colors of \snep\ at wavelengths
longer than $\sim$5000\,\AA. Future low-$z$ samples will also allow a
better quantitative determination of the cutoff in decline rate that
separates ``good standardizable \snep'' from other SNe.

\section{Conclusions}\label{s:conclude}

Using a sample that is larger and more diverse than those studied
previously, we have shown that \snep\ can be calibrated using simple
correlations, yielding a tight Hubble diagram with a dispersion of
$\sim$10\% in distance, not much worse than for SNe\,Ia, if one carefully
constructs the sample of SNe. We have shown that the N06
standardization method, a derivative of the HP02 method, might be
simplified even further because the correction for dust extinction,
based on the current sample, has a low statistical impact on the
scatter in the Hubble diagram.  This implies that a distance
measurement could, in principle, be obtained with a single spectrum of
a SN during the plateau phase, combined with rest-frame $V$ or
$I$-band photometry at a similar time. However, additional (lower S/N)
photometry is required during the plateau phase, in order to reject
non-II-P SNe.

If dust correction is applied, the best-fit solution prefers a very
steep dust law, with $R_V<2$, as recently indicated for SNe\,Ia.
This may have significant implications for SN\,Ia cosmology, where
systematic uncertainties currently dominate. It is further supported
by two pairs of SNe that occurred in the same host galaxies, and whose
distances agree only for low $R_V$ values.

The Hubble diagram in Figure\ \ref{f:HD} is dominated by nearby
objects, where systemic velocities of the host galaxies govern the
error bars.  More exact values of the parameters, and more confidence
in the method, will require the analysis of a sample of SNe in the
Hubble flow.  The most useful redshift range to calibrate the methods
is $z \approx 0.05$--0.1, a distance that has been almost inaccessible
until recently due to the requirement of a very large, and moderately
deep, search. Pan-STARRS \citep{kaiser02} and the Palomar Transient
Factory (PTF; Rau et al. 2009, in prep.) will supply
more \snep\ than any follow-up program could realistically handle. If
subsequent follow-up resources are allocated, those SNe will become
the backbone of any cosmological use of \snep, anchoring the
correlation on more secure grounds.

\acknowledgments 

D.P. wishes to thank A. Gal-Yam, D. Maoz, E. Ofek, and A. Sternberg,
for perpetual useful advice.  D.P. and N.B. were partially supported
by US Department of Energy SciDAC grant DE-FC02-06ER41453. N.B. acknowledges 
NASA support through the GLAST Fellowship Program, NASA 
Cooperative Agreement: NNG06DO90A. The
supernova photometry used here was obtained with KAIT; its
construction and ongoing operation were made possible by donations
from Sun Microsystems, Inc., the Hewlett-Packard Company, AutoScope
Corporation, Lick Observatory, the US National Science Foundation, the
University of California, the Sylvia \& Jim Katzman Foundation, and
the TABASGO Foundation.  Most of the spectra used here were obtained
by A.V.F.'s group with the 3~m Shane reflector at Lick Observatory. We
thank the Lick staff for their dedicated help, as well as the
following for their assistance with some of the observations:
C. Anderson, A. Coil, W. de Vries, B. Grigsby, T. Lowe, M. A. Malkan,
T.  Matheson, M. Papenkova, S. Park, J. Rex, K. Shimasaki, T. Treu,
W. van Breugel, D. Weisz, and D. Winslow.  Some additional spectra
were obtained at the W.~M. Keck Observatory, which is operated as a
scientific partnership among the California Institute of Technology,
the University of California, and the National Aeronautics and Space
Administration; the observatory was made possible by the generous
financial support of the W.~M. Keck Foundation.  We thank
C. V. Griffith and N. Lee for their help in improving and maintaining
the SNDB.  A.V.F.'s supernova group is supported by NSF grant
AST-0607485, US Department of Energy grant DE-FG02-08ER41563, Gary and
Cynthia Bengier, and the TABASGO Foundation.  J.S.B.'s group is
partially supported by NASA/{\it Swift} grant \#NNG05GF55G and a
Hellman Faculty Award.  A.A.M. is supported by a UC Berkeley
Chancellor's Fellowship.  M.M. is grateful for a Postdoctoral
Fellowship from the Miller Institute for Basic Research in Science.
P.E.N. acknowledges support from the US Department of Energy
Scientific Discovery through Advanced Computing program under contract
DE-FG02-06ER06-04. This research used resources of the National Energy
Research Scientific Computing Center, which is supported by the Office
of Science of the US Department of Energy under contract
DE-AC03-76SF00098; we thank them for a generous allocation of
computing time.


\begin{thebibliography}{95}
\expandafter\ifx\csname natexlab\endcsname\relax\def\natexlab#1{#1}\fi

\bibitem[{{Aazami} \& {Li}(2000)}]{aazami00}
{Aazami}, A.~B., \& {Li}, W.~D. 2000, \iaucirc, 7490, 1

\bibitem[{{Alard} \& {Lupton}(1998)}]{alard98}
{Alard}, C., \& {Lupton}, R.~H. 1998, \apj, 503, 325

\bibitem[{{Astier} {et~al.}(2006){Astier}, {Guy}, {Regnault}, {Pain},
  {Aubourg}, {Balam}, {Basa}, {Carlberg}, {Fabbro}, {Fouchez}, {Hook},
  {Howell}, {Lafoux}, {Neill}, {Palanque-Delabrouille}, {Perrett}, {Pritchet},
  {Rich}, {Sullivan}, {Taillet}, {Aldering}, {Antilogus}, {Arsenijevic},
  {Balland}, {Baumont}, {Bronder}, {Courtois}, {Ellis}, {Filiol}, {Gon{\c
  c}alves}, {Goobar}, {Guide}, {Hardin}, {Lusset}, {Lidman}, {McMahon},
  {Mouchet}, {Mourao}, {Perlmutter}, {Ripoche}, {Tao}, \& {Walton}}]{astier06}
{Astier}, P., {et~al.} 2006, \aap, 447, 31

\bibitem[{{Baade}(1926)}]{baade26}
{Baade}, W. 1926, Astronomische Nachrichten, 228, 359

\bibitem[{{Barbon} {et~al.}(1979){Barbon}, {Ciatti}, \& {Rosino}}]{barbon79}
{Barbon}, R., {Ciatti}, F., \& {Rosino}, L. 1979, \aap, 72, 287

\bibitem[{{Barlow} {et~al.}(2005){Barlow}, {Sugerman}, {Fabbri}, {Meixner},
  {Fisher}, {Bowey}, {Panagia}, {Ercolano}, {Clayton}, {Cohen}, {Gledhill},
  {Gordon}, {Tielens}, \& {Zijlstra}}]{barlow05}
{Barlow}, M.~J., {et~al.} 2005, \apjl, 627, L113

\bibitem[{Baron {et~al.}(2004)Baron, Nugent, Branch, \& Hauschildt}]{baron04}
Baron, E., Nugent, P.~E., Branch, D., \& Hauschildt, P.~H. 2004, ApJ, 616, L91

\bibitem[{Baron {et~al.}(2003)Baron, Nugent, Branch, Hauschildt, Turatto, \&
  Cappellaro}]{baron03}
Baron, E., Nugent, P.~E., Branch, D., Hauschildt, P.~H., Turatto, M., \&
  Cappellaro, E. 2003, ApJ, 586, 1199

\bibitem[{{Blondin} \& {Tonry}(2007)}]{blondin07}
{Blondin}, S., \& {Tonry}, J.~L. 2007, \apj, 666, 1024

\bibitem[{{Cardelli} {et~al.}(1989){Cardelli}, {Clayton}, \&
  {Mathis}}]{cardelli89}
{Cardelli}, J.~A., {Clayton}, G.~C., \& {Mathis}, J.~S. 1989, \apj, 345, 245

\bibitem[{Dessart \& Hillier(2006)}]{dessart06}
Dessart, L., \& Hillier, D.~J. 2006, A\&A, 447, 691

\bibitem[{Dessart {et~al.}(2008)Dessart, Blondin, Brown, Hicken, Hillier,
  Holland, Immler, Kirshner, Milne, Modjaz, \& Roming}]{dessart08}
Dessart, L., {et~al.} 2008, ApJ, 675, 644

\bibitem[{{Doggett} \& {Branch}(1985)}]{doggett85}
{Doggett}, J.~B., \& {Branch}, D. 1985, \aj, 90, 2303

\bibitem[{{Draine}(2003)}]{draine03}
Draine, B.~T. 2003, \araa, 41, 241

\bibitem[{Eastman {et~al.}(1996)Eastman, Schmidt, \& Kirshner}]{eastman96}
Eastman, R.~G., Schmidt, B.~P., \& Kirshner, R. 1996, ApJ, 466, 911

\bibitem[{{Elias-Rosa} {et~al.}(2006){Elias-Rosa}, {Benetti}, {Cappellaro},
  {Turatto}, {Mazzali}, {Patat}, {Meikle}, {Stehle}, {Pastorello}, {Pignata},
  {Kotak}, {Harutyunyan}, {Altavilla}, {Navasardyan}, {Qiu}, {Salvo}, \&
  {Hillebrandt}}]{elias-rosa06}
{Elias-Rosa}, N., {et~al.} 2006, \mnras, 369, 1880

\bibitem[{{Elias-Rosa} {et~al.}(2008){Elias-Rosa}, {Benetti}, {Turatto},
  {Cappellaro}, {Valenti}, {Arkharov}, {Beckman}, {di Paola}, {Dolci},
  {Filippenko}, {Foley}, {Krisciunas}, {Larionov}, {Li}, {Meikle},
  {Pastorello}, {Valentini}, \& {Hillebrandt}}]{elias-rosa08}
---. 2008, \mnras, 384, 107

\bibitem[{Ellis {et~al.}(2008)Ellis, Sullivan, Nugent, Howell, Gal-Yam, Astier,
  Balam, Balland, Basa, Carlberg, Conley, Fouchez, Guy, Hardin, Hook, Pain,
  Perrett, Pritchet, \& Regnault}]{ellis08}
Ellis, R.~S., {et~al.} 2008, ApJ, 674, 51

\bibitem[{{Faber} {et~al.}(2003){Faber}, {Phillips}, {Kibrick}, {Alcott},
  {Allen}, {Burrous}, {Cantrall}, {Clarke}, {Coil}, {Cowley}, {Davis}, {Deich},
  {Dietsch}, {Gilmore}, {Harper}, {Hilyard}, {Lewis}, {McVeigh}, {Newman},
  {Osborne}, {Schiavon}, {Stover}, {Tucker}, {Wallace}, {Wei}, {Wirth}, \&
  {Wright}}]{faber03}
{Faber}, S.~M., {et~al.} 2003, in SPIE Proc., ed. M.~{Iye} \& A.~F.~M.
  {Moorwood}, Vol. 4841, 1657

\bibitem[{{Filippenko}(1982)}]{filippenko82}
{Filippenko}, A.~V. 1982, \pasp, 94, 715

\bibitem[{{Filippenko}(1997)}]{filippenko97}
---. 1997, \araa, 35, 309

\bibitem[{{Filippenko}(2005{\natexlab{a}})}]{filippenko05b}
{Filippenko}, A.~V. 2005{\natexlab{a}}, in The Fate of the Most Massive Stars, 
  ed. R.~{Humphreys} \& K.~{Stanek} (San Francisco: ASP, Conf. Ser. Vol. 332), 33

\bibitem[{{Filippenko}(2005{\natexlab{b}})}]{filippenko05}
{Filippenko}, A.~V. 2005{\natexlab{b}}, in White Dwarfs: Cosmological and 
  Galactic Probes, ed. E.~M. {Sion}, S.~{Vennes}, \& H.~L. {Shipman}
  (Dordrecht: Springer), 97

\bibitem[{{Filippenko} {et~al.}(2001){Filippenko}, {Li}, {Treffers}, \&
  {Modjaz}}]{filippenko01}
{Filippenko}, A.~V., {Li}, W.~D., {Treffers}, R.~R., \& {Modjaz}, M. 2001, in
  Small Telescope Astronomy on Global
  Scales, ed. B.~{Paczy\'{n}ski}, W.-P. {Chen}, \& C.~{Lemme} (San Francisco:
  ASP, Conf. Ser. Vol. 246), 121

\bibitem[{{Foley} {et~al.}(2008){Foley}, {Filippenko}, {Aguilera}, {Becker},
  {Blondin}, {Challis}, {Clocchiatti}, {Covarrubias}, {Davis}, {Garnavich},
  {Jha}, {Kirshner}, {Krisciunas}, {Leibundgut}, {Li}, {Matheson}, {Miceli},
  {Miknaitis}, {Pignata}, {Rest}, {Riess}, {Schmidt}, {Smith}, {Sollerman},
  {Spyromilio}, {Stubbs}, {Suntzeff}, {Tonry}, {Wood-Vasey}, \&
  {Zenteno}}]{foley08}
{Foley}, R.~J., {et~al.} 2008, ApJ, 684, 68

\bibitem[{{Foley} {et~al.}(2003){Foley}, {Papenkova}, {Swift}, {Filippenko},
  {Li}, {Mazzali}, {Chornock}, {Leonard}, \& {Van Dyk}}]{foley03}
---. 2003, \pasp, 115, 1220

\bibitem[{{Gal-Yam} {et~al.}(2007){Gal-Yam}, {Cenko}, {Fox}, {Leonard}, {Moon},
  {Sand}, \& {Soderberg}}]{gal-yam07cccp}
{Gal-Yam}, A., {Cenko}, S.~B., {Fox}, D.~B., {Leonard}, D.~C., {Moon}, D.-S.,
  {Sand}, D.~J., \& {Soderberg}, A.~M. 2007, in The
  Multicoloured Landscape of Compact Objects and Their Explosive Origins, ed.
  T.~{di Salvo}, et al. (New York: AIP, Conf. 924), 297

\bibitem[{{Ganeshalingam} {et~al.}(2001){Ganeshalingam}, {Modjaz}, \&
  {Li}}]{ganeshalingam01}
{Ganeshalingam}, M., {Modjaz}, M., \& {Li}, W.~D. 2001, \iaucirc, 7655, 1

\bibitem[{{Gezari} {et~al.}(2008){Gezari}, {Halpern}, {Grupe}, {Yuan},
  {Quimby}, {McKay}, {Chamarro}, {Sisson}, {Akerlof}, {Wheeler}, {Brown},
  {Cenko}, {Rau}, {Djordjevic}, \& {Terndrup}}]{gezari08}
{Gezari}, S., {et~al.} 2008, in press (ArXiv:0808.2812)

\bibitem[{{Goobar}(2008)}]{goobar08}
{Goobar}, A. 2008, submitted (ArXiv:0809.1094)

\bibitem[{{Goobar} \& {Perlmutter}(1995)}]{goobar95}
{Goobar}, A., \& {Perlmutter}, S. 1995, \apj, 450, 14

\bibitem[{Hamuy(2005)}]{hamuy05}
Hamuy, M. 2005, in Cosmic Explosions, ed. J.-M. {Marcaide} \& K.~W. 
{Weiler} (Berlin: Springer-Verlag, IAU Col. 192), 535

\bibitem[{Hamuy \& Pinto(2002)}]{hamuy02}
Hamuy, M., \& Pinto, P.~A. 2002, ApJ, 566, L63

\bibitem[{{Hamuy} {et~al.}(2001){Hamuy}, {Pinto}, {Maza}, {Suntzeff},
  {Phillips}, {Eastman}, {Smith}, {Corbally}, {Burstein}, {Li}, {Ivanov},
  {Moro-Martin}, {Strolger}, {de Souza}, {dos Anjos}, {Green}, {Pickering},
  {Gonz{\'a}lez}, {Antezana}, {Wischnjewsky}, {Galaz}, {Roth}, {Persson}, \&
  {Schommer}}]{hamuy01}
{Hamuy}, M., {et~al.} 2001, \apj, 558, 615

\bibitem[{{Horne}(1986)}]{horne86}
{Horne}, K. 1986, \pasp, 98, 609

\bibitem[{{Jiang} \& {Qiu}(2001)}]{jiang01}
{Jiang}, X.~J., \& {Qiu}, Y.~L. 2001, \iaucirc, 7641, 3

\bibitem[{{Kaiser} {et~al.}(2002){Kaiser}, {Aussel}, {Burke}, {Boesgaard},
  {Chambers}, {Chun}, {Heasley}, {Hodapp}, {Hunt}, {Jedicke}, {Jewitt},
  {Kudritzki}, {Luppino}, {Maberry}, {Magnier}, {Monet}, {Onaka}, {Pickles},
  {Rhoads}, {Simon}, {Szalay}, {Szapudi}, {Tholen}, {Tonry}, {Waterson}, \&
  {Wick}}]{kaiser02}
{Kaiser}, N., {et~al.} 2002, in SPIE Proc., ed. J.~A. {Tyson} \& S.~{Wolff},
  Vol. 4836, 154

\bibitem[{{Kelly}(2007)}]{kelly07}
{Kelly}, B.~C. 2007, \apj, 665, 1489

\bibitem[{Kirshner \& Kwan(1975)}]{kirshner75}
Kirshner, R.~P., \& Kwan, J. 1975, ApJ, 197, 415

\bibitem[{{Kloehr} {et~al.}(2005){Kloehr}, {Muendlein}, {Li}, {Yamaoka}, \&
  {Itagaki}}]{kloehr05}
{Kloehr}, W., {Muendlein}, R., {Li}, W., {Yamaoka}, H., \& {Itagaki}, K. 2005,
  \iaucirc, 8553, 1

\bibitem[{{Klotz} {et~al.}(2002){Klotz}, {Puckett}, {Langoussis}, {Wood-Vasey},
  {Aldering}, {Nugent}, \& {Stephens}}]{klotz02}
{Klotz}, A., {Puckett}, T., {Langoussis}, A., {Wood-Vasey}, W.~M., {Aldering},
  G., {Nugent}, P., \& {Stephens}, R. 2002, \iaucirc, 7986, 1

\bibitem[{{Kowalski} {et~al.}(2008){Kowalski}, {Rubin}, {Aldering},
  {Agostinho}, {Amadon}, {Amanullah}, {Balland}, {Barbary}, {Blanc}, {Challis},
  {Conley}, {Connolly}, {Covarrubias}, {Dawson}, {Deustua}, {Ellis}, {Fabbro},
  {Fadeyev}, {Fan}, {Farris}, {Folatelli}, {Frye}, {Garavini}, {Gates},
  {Germany}, {Goldhaber}, {Goldman}, {Goobar}, {Groom}, {Haissinski}, {Hardin},
  {Hook}, {Kent}, {Kim}, {Knop}, {Lidman}, {Linder}, {Mendez}, {Meyers},
  {Miller}, {Moniez}, {Mourao}, {Newberg}, {Nobili}, {Nugent}, {Pain},
  {Perdereau}, {Perlmutter}, {Phillips}, {Prasad}, {Quimby}, {Regnault},
  {Rich}, {Rubenstein}, {Ruiz-Lapuente}, {Santos}, {Schaefer}, {Schommer},
  {Smith}, {Soderberg}, {Spadafora}, {Strolger}, {Strovink}, {Suntzeff},
  {Suzuki}, {Thomas}, {Walton}, {Wang}, {Wood-Vasey}, \& {Yun}}]{kowalski08}
{Kowalski}, M., {et~al.} 2008, in press (ArXiv:0804.4142)

\bibitem[{{Krisciunas} {et~al.}(2007){Krisciunas}, {Garnavich}, {Stanishev},
  {Suntzeff}, {Prieto}, {Espinoza}, {Gonzalez}, {Salvo}, {Elias de la Rosa},
  {Smartt}, {Maund}, \& {Kudritzki}}]{krisciunas07}
{Krisciunas}, K., {et~al.} 2007, \aj, 133, 58

\bibitem[{{Landolt}(1992)}]{landolt92}
{Landolt}, A.~U. 1992, \aj, 104, 340

\bibitem[{Leonard {et~al.}(2003)Leonard, Kanbur, Ngeow, \& Tanvir}]{leonard03}
Leonard, D.~C., Kanbur, S.~M., Ngeow, C.~C., \& Tanvir, N.~R. 2003, ApJ, 594,
  247

\bibitem[{{Leonard} {et~al.}(2002{\natexlab{a}}){Leonard}, {Filippenko},
  {Gates}, {Li}, {Eastman}, {Barth}, {Bus}, {Chornock}, {Coil}, {Frink},
  {Grady}, {Harris}, {Malkan}, {Matheson}, {Quirrenbach}, \&
  {Treffers}}]{leonard02em}
{Leonard}, D.~C., {et~al.} 2002{\natexlab{a}}, \pasp, 114, 35

\bibitem[{{Leonard} {et~al.}(2002{\natexlab{b}}){Leonard}, {Filippenko}, {Li},
  {Matheson}, {Kirshner}, {Chornock}, {Van Dyk}, {Berlind}, {Calkins},
  {Challis}, {Garnavich}, {Jha}, \& {Mahdavi}}]{leonard02gi}
---. 2002{\natexlab{b}}, \aj, 124, 2490


\bibitem[{{Li}(1999{\natexlab{a}})}]{li99a}
{Li}, W. 1999{\natexlab{a}}, \iaucirc, 7135, 1

\bibitem[{{Li}(1999{\natexlab{b}})}]{li99}
---. 1999{\natexlab{b}}, \iaucirc, 7294, 1

\bibitem[{{Li}(2002)}]{li02}
---. 2002, \iaucirc, 8005, 1

\bibitem[{{Li}(2003)}]{li03}
---. 2003, \iaucirc, 8184, 1

\bibitem[{{Li} {et~al.}(2001{\natexlab{a}}){Li}, {Fan}, {Qiu}, {Hu}, \&
  {Schwartz}}]{li01a}
{Li}, W., {Fan}, Y., {Qiu}, Y.~L., {Hu}, J.~Y., \& {Schwartz}, M.
  2001{\natexlab{a}}, \iaucirc, 7591, 1

\bibitem[{{Li} {et~al.}(2001{\natexlab{b}}){Li}, {Filippenko}, {Gates},
  {Chornock}, {Gal-Yam}, {Ofek}, {Leonard}, {Modjaz}, {Rich}, {Riess}, \&
  {Treffers}}]{li01cx}
{Li}, W., {et~al.} 2001{\natexlab{b}}, \pasp, 113, 1178

\bibitem[{Li {et~al.}(2007)Li, Wang, Dyk, Cuillandre, Foley, \&
  Filippenko}]{li07}
Li, W., Wang, X., Dyk, S. D.~V., Cuillandre, J.-C., Foley, R.~J., \&
  Filippenko, A.~V. 2007, ApJ, 661, 1013

\bibitem[{{Llapasset}(2003)}]{llapasset03}
{Llapasset}, J. 2003, \iaucirc, 8219, 2

\bibitem[{{Matheson} {et~al.}(2000){Matheson}, {Filippenko}, {Ho}, {Barth}, \&
  {Leonard}}]{matheson00}
{Matheson}, T., {Filippenko}, A.~V., {Ho}, L.~C., {Barth}, A.~J., \& {Leonard},
  D.~C. 2000, \aj, 120, 1499

\bibitem[{{Miller} {et~al.}(2008){Miller}, {Chornock}, {Perley},
  {Ganeshalingam}, {Li}, {Butler}, {Bloom}, {Smith}, {Modjaz}, {Poznanski},
  {Filippenko}, {Griffith}, {Shiode}, \& {Silverman}}]{miller08}
{Miller}, A.~A., {et~al.} 2008, in press (ArXiv:0808.2193)

\bibitem[{Miller \& Stone(1993)}]{miller93}
Miller, J.~S., \& Stone, R. P.~S. 1993, Lick Obs. Tech. Rep. 66,
  (Santa Cruz: Lick Obs.)

\bibitem[{{Modjaz} \& {Li}(2001)}]{modjaz01b}
{Modjaz}, M., \& {Li}, W.~D. 2001, \iaucirc, 7682, 1

\bibitem[{{Modjaz} {et~al.}(2001){Modjaz}, {Li}, {Filippenko}, {King},
  {Leonard}, {Matheson}, {Treffers}, \& {Riess}}]{modjaz01}
{Modjaz}, M., {Li}, W., {Filippenko}, A.~V., {King}, J.~Y., {Leonard}, D.~C.,
  {Matheson}, T., {Treffers}, R.~R., \& {Riess}, A.~G. 2001, \pasp, 113, 308

\bibitem[{{Moro} \& {Munari}(2000)}]{moro00}
{Moro}, D., \& {Munari}, U. 2000, \aaps, 147, 361

\bibitem[{{Nakano} {et~al.}(2001){Nakano}, {Itagaki}, {Li}, \&
  {Schwartz}}]{nakano01}
{Nakano}, S., {Itagaki}, K., {Li}, W.~D., \& {Schwartz}, M. 2001, \iaucirc,
  7628, 2

\bibitem[{{Nakano} \& {Kushida}(1999)}]{nakano99}
{Nakano}, S., \& {Kushida}, R. 1999, \iaucirc, 7329, 1

\bibitem[{{Nobili} \& {Goobar}(2008)}]{nobili08}
{Nobili}, S., \& {Goobar}, A. 2008, \aap, 487, 19

\bibitem[{Nugent {et~al.}(2006)Nugent, Sullivan, Ellis, Gal-Yam, Leonard,
  Howell, Astier, Carlberg, Conley, Fabbro, Fouchez, Neill, Pain, Perrett,
  Pritchet, \& Regnault}]{nugent06}
Nugent, P., {et~al.} 2006, ApJ, 645, 841

\bibitem[{{Ofek} {et~al.}(2007){Ofek}, {Cameron}, {Kasliwal}, {Gal-Yam}, {Rau},
  {Kulkarni}, {Frail}, {Chandra}, {Cenko}, {Soderberg}, \& {Immler}}]{ofek07}
{Ofek}, E.~O., {et~al.} 2007, \apjl, 659, L13

\bibitem[{{Oke} {et~al.}(1995){Oke}, {Cohen}, {Carr}, {Cromer}, {Dingizian},
  {Harris}, {Labrecque}, {Lucinio}, {Schaal}, {Epps}, \& {Miller}}]{oke95}
{Oke}, J.~B., {et~al.} 1995, \pasp, 107, 375

\bibitem[{{Papenkova} \& {Li}(2000)}]{papenkova00}
{Papenkova}, M., \& {Li}, W.~D. 2000, \iaucirc, 7406, 1

\bibitem[{{Pastorello} {et~al.}(2006){Pastorello}, {Sauer}, {Taubenberger},
  {Mazzali}, {Nomoto}, {Kawabata}, {Benetti}, {Elias-Rosa}, {Harutyunyan},
  {Navasardyan}, {Zampieri}, {Iijima}, {Botticella}, {di Rico}, {Del Principe},
  {Dolci}, {Gagliardi}, {Ragni}, \& {Valentini}}]{pastorello06}
{Pastorello}, A., {et~al.} 2006, \mnras, 370, 1752

\bibitem[{{Phillips} \& {Davis}(1995)}]{phillips95}
{Phillips}, A.~C., \& {Davis}, L.~E. 1995, in Astronomical Data Analysis 
  Software and Systems IV, ed. R.~A. {Shaw}, H.~E. {Payne}, \& J.~J.~E. {Hayes} 
 (San Francisco: ASP, Conf. Ser. Vol. 77), 297

\bibitem[{{Poznanski} {et~al.}(2002){Poznanski}, {Gal-Yam}, {Maoz},
  {Filippenko}, {Leonard}, \& {Matheson}}]{poznanski02}
{Poznanski}, D., {Gal-Yam}, A., {Maoz}, D., {Filippenko}, A.~V., {Leonard},
  D.~C., \& {Matheson}, T. 2002, \pasp, 114, 833

\bibitem[{{Pozzo} {et~al.}(2006){Pozzo}, {Meikle}, {Rayner}, {Joseph},
  {Filippenko}, {Foley}, {Li}, {Mattila}, \& {Sollerman}}]{pozzo06}
{Pozzo}, M., {et~al.} 2006, \mnras, 368, 1169

\bibitem[{{Quimby} {et~al.}(2007){Quimby}, {Aldering}, {Wheeler},
  {H{\"o}flich}, {Akerlof}, \& {Rykoff}}]{quimby07}
{Quimby}, R.~M., {Aldering}, G., {Wheeler}, J.~C., {H{\"o}flich}, P.,
  {Akerlof}, C.~W., \& {Rykoff}, E.~S. 2007, \apjl, 668, L99

\bibitem[{{Rich}(2005)}]{rich05}
{Rich}, D. 2005, \iaucirc, 8500, 2

\bibitem[{{Riess} {et~al.}(2007){Riess}, {Strolger}, {Casertano}, {Ferguson},
  {Mobasher}, {Gold}, {Challis}, {Filippenko}, {Jha}, {Li}, {Tonry}, {Foley},
  {Kirshner}, {Dickinson}, {MacDonald}, {Eisenstein}, {Livio}, {Younger}, {Xu},
  {Dahl{\'e}n}, \& {Stern}}]{riess07}
{Riess}, A.~G., {et~al.} 2007, \apj, 659, 98

\bibitem[{{Schlegel} {et~al.}(1998){Schlegel}, {Finkbeiner}, \&
  {Davis}}]{schlegel98}
{Schlegel}, D.~J., {Finkbeiner}, D.~P., \& {Davis}, M. 1998, \apj, 500, 525

\bibitem[{{Schlegel}(1996)}]{schlegel96}
{Schlegel}, E.~M. 1996, \aj, 111, 1660

\bibitem[{{Schmidt} {et~al.}(1998){Schmidt}, {Suntzeff}, {Phillips},
  {Schommer}, {Clocchiatti}, {Kirshner}, {Garnavich}, {Challis}, {Leibundgut},
  {Spyromilio}, {Riess}, {Filippenko}, {Hamuy}, {Smith}, {Hogan}, {Stubbs},
  {Diercks}, {Reiss}, {Gilliland}, {Tonry}, {Maza}, {Dressler}, {Walsh}, \&
  {Ciardullo}}]{schmidt98}
{Schmidt}, B.~P., {et~al.} 1998, \apj, 507, 46

\bibitem[{{Singer} \& {Li}(2004)}]{singer04}
{Singer}, D., \& {Li}, W. 2004, \iaucirc, 8387, 1

\bibitem[{{Smartt} {et~al.}(2008){Smartt}, {Eldridge}, {Crockett}, \&
  {Maund}}]{smartt08}
{Smartt}, S.~J., {Eldridge}, J.~J., {Crockett}, R.~M., \& {Maund}, J.~R. 2008,
  submitted (ArXiv:0809.0403)

\bibitem[{{Smith} {et~al.}(2008){Smith}, {Chornock}, {Li}, {Ganeshalingam},
  {Silverman}, {Foley}, {Filippenko}, \& {Barth}}]{smith08tf}
{Smith}, N., {Chornock}, R., {Li}, W., {Ganeshalingam}, M., {Silverman}, J.~M.,
  {Foley}, R.~J., {Filippenko}, A.~V., \& {Barth}, A.~J. 2008, \apj, 686,
  467

\bibitem[{{Smith} {et~al.}(2007){Smith}, {Li}, {Foley}, {Wheeler}, {Pooley},
  {Chornock}, {Filippenko}, {Silverman}, {Quimby}, {Bloom}, \&
  {Hansen}}]{smith07}
{Smith}, N., {et~al.} 2007, \apj, 666, 1116

\bibitem[{{Sofue} \& {Rubin}(2001)}]{sofue01}
{Sofue}, Y., \& {Rubin}, V. 2001, \araa, 39, 137

\bibitem[{{Stritzinger} {et~al.}(2002){Stritzinger}, {Hamuy}, {Suntzeff},
  {Smith}, {Phillips}, {Maza}, {Strolger}, {Antezana}, {Gonz{\'a}lez},
  {Wischnjewsky}, {Candia}, {Espinoza}, {Gonz{\'a}lez}, {Stubbs}, {Becker},
  {Rubenstein}, \& {Galaz}}]{stritzinger02}
{Stritzinger}, M., {et~al.} 2002, \aj, 124, 2100

\bibitem[{{Sullivan} {et~al.}(2006){Sullivan}, {Howell}, {Perrett}, {Nugent},
  {Astier}, {Aubourg}, {Balam}, {Basa}, {Carlberg}, {Conley}, {Fabbro},
  {Fouchez}, {Guy}, {Hook}, {Lafoux}, {Neill}, {Pain}, {Palanque-Delabrouille},
  {Pritchet}, {Regnault}, {Rich}, {Taillet}, {Aldering}, {Baumont}, {Bronder},
  {Filiol}, {Knop}, {Perlmutter}, \& {Tao}}]{sullivan06a}
{Sullivan}, M., {et~al.} 2006, \aj, 131, 960

\bibitem[{{Tonry} {et~al.}(2000){Tonry}, {Blakeslee}, {Ajhar}, \&
  {Dressler}}]{tonry00}
{Tonry}, J.~L., {Blakeslee}, J.~P., {Ajhar}, E.~A., \& {Dressler}, A. 2000,
  \apj, 530, 625

\bibitem[{{Tonry} \& {Davis}(1979)}]{tonry79}
{Tonry}, J., \& {Davis}, M. 1979, \aj, 84, 1511

\bibitem[{{Wade} \& {Horne}(1988)}]{wade88}
{Wade}, R.~A., \& {Horne}, K. 1988, \apj, 324, 411

\bibitem[{{Wang}(2005)}]{wang05}
{Wang}, L. 2005, \apjl, 635, L33

\bibitem[{{Wang} {et~al.}(2008){Wang}, {Li}, {Filippenko}, {Krisciunas},
  {Suntzeff}, {Li}, {Zhang}, {Deng}, {Foley}, {Ganeshalingam}, {Li}, {Lou},
  {Qiu}, {Shang}, {Silverman}, {Zhang}, \& {Zhang}}]{wang08}
{Wang}, X., {et~al.} 2008, \apj, 675, 626

\bibitem[{{Weiner} {et~al.}(2005){Weiner}, {Phillips}, {Faber}, {Willmer},
  {Vogt}, {Simard}, {Gebhardt}, {Im}, {Koo}, {Sarajedini}, {Wu}, {Forbes},
  {Gronwall}, {Groth}, {Illingworth}, {Kron}, {Rhodes}, {Szalay}, \&
  {Takamiya}}]{weiner05}
{Weiner}, B.~J., {et~al.} 2005, \apj, 620, 595

\bibitem[{{Wood-Vasey} {et~al.}(2007){Wood-Vasey}, {Miknaitis}, {Stubbs},
  {Jha}, {Riess}, {Garnavich}, {Kirshner}, {Aguilera}, {Becker}, {Blackman},
  {Blondin}, {Challis}, {Clocchiatti}, {Conley}, {Covarrubias}, {Davis},
  {Filippenko}, {Foley}, {Garg}, {Hicken}, {Krisciunas}, {Leibundgut}, {Li},
  {Matheson}, {Miceli}, {Narayan}, {Pignata}, {Prieto}, {Rest}, {Salvo},
  {Schmidt}, {Smith}, {Sollerman}, {Spyromilio}, {Tonry}, {Suntzeff}, \&
  {Zenteno}}]{wood-vasey07}
{Wood-Vasey}, W.~M., {et~al.} 2007, \apj, 666, 694

\bibitem[{{Yu} \& {Li}(2000)}]{yu00}
{Yu}, C., \& {Li}, W.~D. 2000, \iaucirc, 7476, 1

\bibitem[{{Zwitter} {et~al.}(2004){Zwitter}, {Munari}, \&
  {Moretti}}]{zwitter04}
{Zwitter}, T., {Munari}, U., \& {Moretti}, S. 2004, \iaucirc, 8413, 1

\end{thebibliography}

\end{document}